\documentclass[sigconf, screen, authorversion]{acmart}

\usepackage{graphicx}
\usepackage{natbib}
\usepackage{booktabs}
\usepackage{balance}
\usepackage{xcolor}
\usepackage{soul}
\usepackage{subcaption}
\usepackage{multirow}
\usepackage{array}
\usepackage{multirow}

\definecolor{graytext}{gray}{0.5}
\newcommand{\punit}[1]{{\color{graytext}(#1)}}

\copyrightyear{2026}
\acmYear{2026}
\setcopyright{cc}
\setcctype{by}
\acmConference[CUI '26]{ACM Conversational User Interfaces 2026}{July 21--24, 2026}{Bremen, Germany}
\acmBooktitle{ACM Conversational User Interfaces 2026 (CUI '26), July 21--24, 2026, Bremen, Germany}
\acmDOI{10.1145/3816046.3816204}
\acmISBN{979-8-4007-2741-2/2026/07}

\begin{document}

\title{GroupEnvoy: A Conversational Agent Speaking for the Outgroup to Foster Intergroup Relations}

\author{Koken Hata}
\orcid{0009-0001-6053-8537}
\affiliation{%
    \institution{The University of Tokyo}
    \city{Tokyo}
    \country{Japan}
}
\email{hata313@g.ecc.u-tokyo.ac.jp}

\author{Rintaro Chujo}
\orcid{0000-0002-4499-7047}
\affiliation{%
    \institution{The University of Tokyo}
    \city{Tokyo}
    \country{Japan}
}
\email{chujo@nae-lab.org}

\author{Reina Takamatsu}
\orcid{0000-0002-9044-5446}
\affiliation{%
    \institution{Institute of Science Tokyo}
    \city{Tokyo}
    \country{Japan}
}
\email{takamatsu.r.e606@m.isct.ac.jp}

\author{Wenzhen Xu}
\orcid{0000-0003-3969-337X}
\affiliation{%
    \institution{Hitotsubashi University}
    \city{Tokyo}
    \country{Japan}
}
\email{wenzhen.xu@r.hit-u.ac.jp}

\author{Yukino Baba}
\orcid{0000-0001-5310-9841}
\affiliation{%
    \institution{The University of Tokyo}
    \city{Tokyo}
    \country{Japan}
}
\email{yukino-baba@g.ecc.u-tokyo.ac.jp}



\begin{abstract}
    Conversational agents have the potential to support intergroup relations when psychological or linguistic barriers prevent direct interaction. Based on intergroup contact theory, we propose GroupEnvoy, a text-based conversational agent that represents outgroup perspectives during ingroup discussions. Its dialogue is grounded in data from a prior outgroup-only discussion. To evaluate this approach and derive design principles, we conducted a mixed-methods, between-subjects study with university students, in which host-country students formed the ingroup and international students formed the outgroup. Ingroup students performed a collaborative task while engaging with outgroup perspectives, either by interacting with GroupEnvoy (AI-mediated contact) or by reading a static document (passive exposure). Quantitatively, AI-mediated contact demonstrated a directional reduction in intergroup anxiety and an improvement in perspective-taking. Qualitatively, AI-mediated contact enhanced outcome expectancies and directed empathy toward the outgroup's evaluations of the ingroup, whereas passive exposure fostered future contact intentions and elicited empathy toward the outgroup's lived experiences. These findings present AI-mediated contact as a promising paradigm for improving intergroup relations.
\end{abstract}

\begin{teaserfigure}
  \centering
  \includegraphics[width=\textwidth]{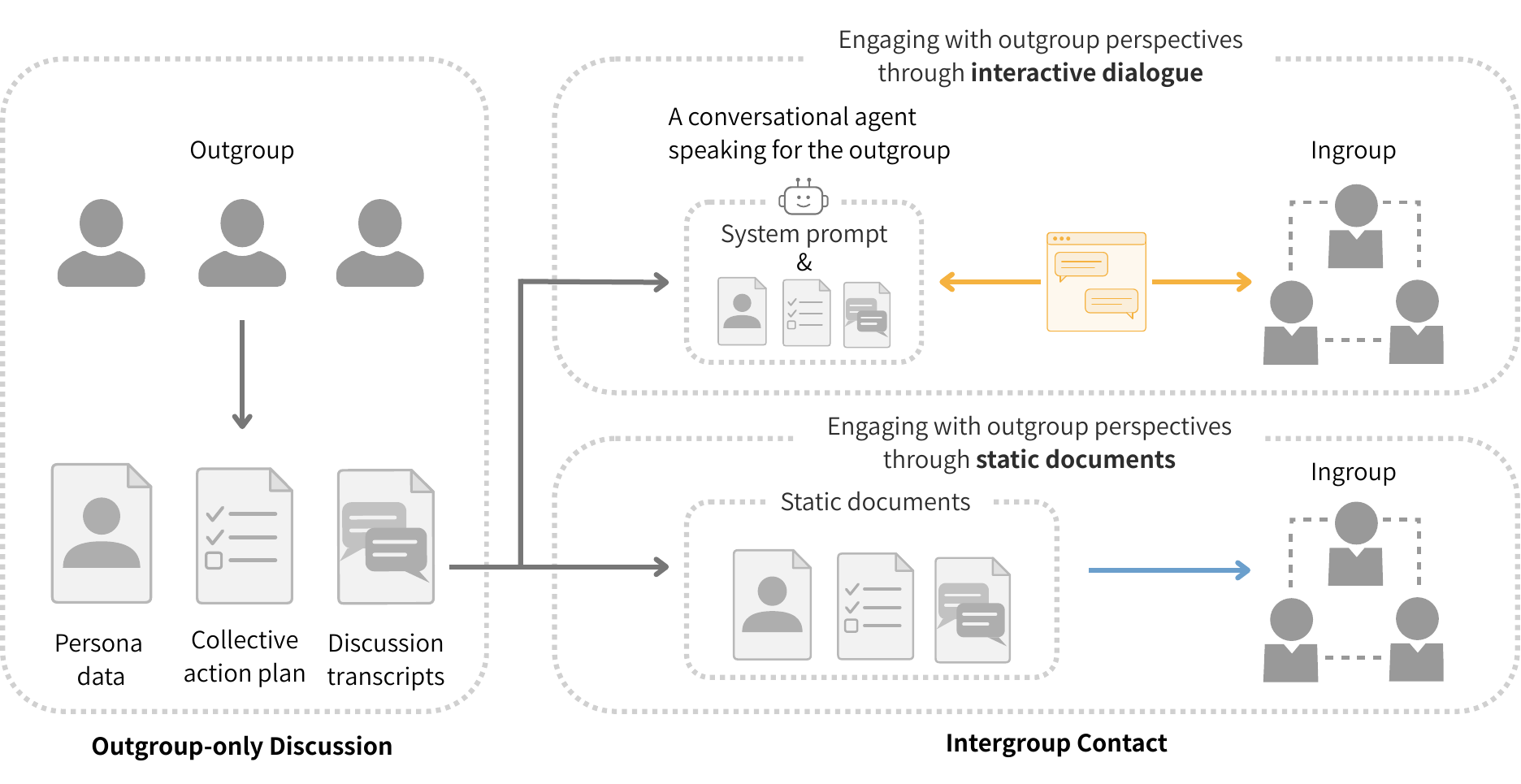}
  \caption{We introduce GroupEnvoy, a text-based conversational agent that facilitates intergroup contact by representing outgroup perspectives to ingroup members. Its dialogue is grounded in outgroup discussion data.}
  \Description{A conceptual diagram with two parts. The left part, labeled Outgroup-only Discussion, shows three outgroup members at the top, with an arrow pointing down to three document icons: persona data, a collective action plan, and discussion transcripts. The right part, labeled Intergroup Contact, is split into two rows. The upper row, titled Engaging with outgroup perspectives through AI-mediated dialogue, shows GroupEnvoy, a conversational agent speaking for the outgroup, exchanging text-chat messages with three ingroup members. The lower row, titled Engaging with outgroup perspectives through static documents, shows the same three documents being read by three ingroup members. Arrows from the left part feed into both rows of the right part.}
  \label{fig:teaser}
\end{teaserfigure}

\maketitle

\section{Introduction}

As globalization accelerates and migration continues to rise, societies worldwide are characterized by increasing contact among groups with diverse cultural, ethnic, and national backgrounds \cite{dovidio_reducing_2017}. However, increased opportunities for contact do not necessarily lead to interaction. Intergroup anxiety and prejudice often lead people to avoid cross-group interactions, maintain social distance, and sometimes contribute to intergroup tensions \cite{paolini_intergroup_2021, al_ramiah_intergroup_2013}.

Intergroup Contact Theory \cite{allport_nature_1954, pettigrew_intergroup_1998} suggests that direct contact between members of different groups can reduce prejudice when four conditions are met: \textit{equal status}, \textit{common goals}, \textit{cooperation}, and \textit{institutional support}.
This model has accumulated extensive empirical support across diverse groups and contexts \cite{pettigrew_meta-analytic_2006}, informing integration policies and reconciliation programs worldwide \cite{paolini_intergroup_2021}.

However, direct intergroup contact often remains difficult to achieve. Psychological, linguistic, and geographic barriers render direct interactions aversive or inaccessible, and contact under unfavorable conditions risks reinforcing negative impressions \cite{amichai-hamburger_contact_2006, white_text-based_2020, crabtree_can_2025}. These limitations have motivated electronic contact (E-contact) \cite{white_improving_2015}, which leverages advances in digital communication technologies to facilitate online text-based or voice-based intergroup interactions.

More recently, the emergence of large language models (LLMs) has enabled more flexible AI-powered approaches to E-contact. \citet{sahab_e-contact_2024} demonstrated that an AI facilitator that leveraged argument-mining techniques in synchronous interethnic discussions significantly reduced intergroup anxiety and prejudice towards other ethnic groups in Afghanistan. In the context of consensus-building, approaches in which groups exchange views with an LLM-powered AI agent rather than interacting directly have also shown promise for identifying common ground \cite{tessler_ai_2024, konya_using_2025}. In both cases, the agent's non-judgmental neutrality alleviates concerns about facilitator bias and social evaluation anxiety, enabling a fairer communication process \cite{sahab_e-contact_2024, tessler_ai_2024, hermann_reducing_2025}.

Nevertheless, each approach faces important limitations. Intergroup contact facilitated by an agent still presupposes a shared language, and even if AI translation bridges language barriers, intergroup anxiety persists because it stems from the anticipation of negative evaluation by outgroup members \cite{stephan_intergroup_1985}. Under such anxiety, individuals rely on schematic processing characterized by a narrowed attention to stereotype-consistent information and the discounting of counterstereotypic behaviors \cite{stephan_intergroup_1985}. Consequently, rather than fostering positive schema changes, such interactions often fail to facilitate meaningful relationship-building. Consensus-building approaches instead bypass direct intergroup communication altogether and place little emphasis on activating the mediating mechanisms crucial for enhancing intergroup relations, with each group exchanging views in isolation rather than engaging with outgroup members. 

To overcome these limitations, we introduce AI-mediated contact, a novel paradigm implemented in \emph{GroupEnvoy}, a text-based conversational agent that conveys outgroup perspectives to ingroup members. Grounded in data from a prior outgroup-only discussion, GroupEnvoy enables ingroup members to engage in interactive, text-based dialogue that reflects outgroup perspectives during group discussions. This approach is designed to reduce the intergroup anxiety inherent in synchronous intergroup communications while preserving outgroup perspectives in their own words. More broadly, this framework aims to mitigate psychological barriers to intergroup contact and foster future direct contact.

Our design is motivated by evidence from E-contact research showing that interactive contact (e.g., text- or voice-based discussion) with outgroup members reduces prejudice more effectively than other indirect forms of intergroup contact \cite{pereira_da_costa_does_2024}.
This suggests that the experience of interactive conversation between ingroups and outgroups, regardless of its specific form, is effective in improving intergroup relationships. Such effects are understood to operate through intergroup anxiety and increased affective empathy and perspective-taking \cite{pettigrew_how_2008}. Additionally, in text-based E-contact, decreased intergroup anxiety was mediated by enhanced outcome expectancies afforded by greater control over self-presentation \cite{white_improving_2019}. 

Building on these findings, GroupEnvoy offers text-based, synchronous contact with outgroup perspectives while avoiding the social evaluation anxiety and language barriers associated with direct intergroup interaction. We hypothesize that AI-mediated contact through GroupEnvoy can improve outgroup attitudes and reduce intergroup anxiety, thereby promoting future contact intentions to engage directly with outgroup members. We also seek to understand which design choices in AI-mediated contact reduce psychological barriers, enabling the approach to be applied in other settings. We therefore address the following research questions:

\begin{quote}
    \textbf{RQ1:} Does AI-mediated contact through GroupEnvoy promote ingroup members' positive outgroup attitudes, positive outcome expectancies, and future contact intentions while reducing intergroup anxiety?
\end{quote}
\begin{quote}
    \textbf{RQ2:} What design principles for AI-mediated contact can help lower psychological barriers to direct intergroup contact and promote future direct interactions?
\end{quote}

To evaluate this framework, we applied GroupEnvoy to the context of intergroup relations between Japanese domestic students (ingroup) and Chinese international students (outgroup). In addition to language barriers, interactions between these two groups are often marked by subtle friction, as differing cultural norms can lead to misunderstandings or reluctance to communicate \cite{tajima_educational_2025, sakakibara_intercultural_2017}. We conducted a between-subjects experiment with two conditions. In the experimental condition, ingroup participants collaborated via an online meeting tool and interacted with GroupEnvoy through a text-based chat interface to pursue a shared goal with the outgroup. In the control condition, participants received the same outgroup data as a static document. Because both conditions received the same outgroup data, this design isolates the effect of interactive contact from that of mere exposure to outgroup information.

When combining pre-post psychological measures with template analysis of post-experiment interviews, significant main effects of Time were observed across most measures, indicating that exposure to outgroup perspectives promoted attitude change in both conditions. Although no statistically significant Group $\times$ Time interaction was observed, medium-to-large effect sizes for intergroup anxiety and perspective-taking suggest that AI-mediated contact offers additional benefits beyond passive text exposure. Qualitative analysis demonstrated that interactive dialogue with GroupEnvoy activated distinct psychological mechanisms unique to the experimental condition. Participants in the experimental condition exhibited enhanced outcome expectancies, engagement with \textit{meta-perceptions} of the outgroup's perception of the ingroup, and three distinct patterns of perspective-taking: sustained engagement, delegation to GroupEnvoy, and progressive shifts toward consensus-building. However, these findings also highlight several limitations. GroupEnvoy's facilitative tendencies risked substituting rather than scaffolding users' perspective-taking effort, and its collective representation of the outgroup made it difficult for participants to perceive the individual people behind the group identity. The main contributions of this work are as follows:
\begin{enumerate}
    \item We introduce AI-mediated contact as a novel paradigm for indirect intergroup contact, in which a conversational agent grounded in outgroup discussion data conveys outgroup perspectives during ingroup discussions, and present GroupEnvoy as its implementation.
    \item We provide empirical mixed-methods insights into the potential for AI-mediated contact to reduce intergroup anxiety and promote perspective-taking, compared with passive exposure to the same outgroup information.
    \item We derive refined design principles for AI-mediated intergroup contact by evaluating GroupEnvoy’s performance. Our analysis identifies effective features and inherent tensions that future implementations should address.
\end{enumerate}

\section{Related Works}

\subsection{Intergroup Contact Theory and E-contact} \label{sec:contact_theory}

Intergroup contact theory posits that direct contact between members of different groups reduces prejudice when four conditions are present: \textit{equal status}, \textit{common goals}, \textit{cooperation}, and \textit{institutional support} \cite{allport_nature_1954, pettigrew_intergroup_1998}. \citet{pettigrew_meta-analytic_2006} confirmed this pattern in a large-scale meta-analysis, showing that contact typically reduces prejudice across diverse groups and contexts. The four conditions were found to be not essential but rather facilitating conditions that promote effective outcomes. However, direct intergroup contact remains difficult to achieve due to psychological, linguistic, or geographic barriers \cite{amichai-hamburger_contact_2006, white_text-based_2020}. This motivates the development of indirect contact strategies to improve intergroup relations without direct interaction.

E-contact, defined as ``computer-mediated contact involving an engagement of self in the intergroup relationship'' \cite{white_dual_2012}, leverages advances in social networking and information and communication technology. Current research primarily distinguishes two forms: \textit{direct E-contact} via synchronous text or voice chat with outgroup members \cite{white_dual_2012, kim_intergroup_2018}, and \textit{embodied contact} via first-person inhabitation of outgroup avatars in virtual reality \cite{banakou_virtual_2016}. Multiple meta-analyses have evaluated these strategies \cite{imperato_allport_2021, pereira_da_costa_does_2024, white_text-based_2020}, and \citet{pereira_da_costa_does_2024} found that direct E-contact yields significantly larger effects than other indirect contact methods (e.g., extended contact \cite{wright_extended_1997}), whereas embodied contact alone fails to reach statistical significance. The authors attribute this disparity to the degree of interactive engagement with real or perceived-as-real outgroup members, identifying it as the key driver of E-contact effectiveness.


Despite its relative effectiveness, direct E-contact requires bilateral participation and a shared language, limiting its feasibility across linguistically distant groups. Even when a shared language exists, direct outgroup interaction can heighten intergroup anxiety by activating concerns about social evaluation \cite{stephan_intergroup_1985}. Furthermore, while immersive alternatives like embodied contact exist, they often yield nonsignificant results despite high implementation costs \cite{pereira_da_costa_does_2024}. 

Recent LLM-powered approaches partially address these barriers. \citet{sahab_e-contact_2024} deployed a conversational facilitator in synchronous discussions across ethnic groups, significantly reducing prejudice and intergroup anxiety. However, this bilateral design still presupposes a shared language, leaving groups separated by linguistic boundaries behind. GroupEnvoy addresses this gap by representing outgroup perspectives through ingroup members' language and grounding its responses in prior outgroup-only discussions. This design preserves the interactive engagement identified as the primary driver of E-contact effectiveness \cite{pereira_da_costa_does_2024}, thereby balancing broad applicability with cost-effectiveness for linguistically distant groups.

\subsection{Mediating Mechanisms of Intergroup Contact} \label{sec:mediating_mechanisms}

\citet{pettigrew_how_2008} identified three key mediators of improved intergroup relations: increased knowledge of the outgroup, reduced intergroup anxiety, and the promotion of affective empathy and perspective-taking. Their findings highlighted that intergroup anxiety and affective empathy, alongside perspective-taking, exerted particularly significant effects on prejudice reduction. Among these mechanisms, intergroup anxiety is defined as the psychological state that emerges when individuals anticipate negative outcomes from future direct contact \cite{stephan_intergroup_1985}. This construct has been consistently identified as a key mediator in reducing prejudice and improving outgroup attitudes \cite{pettigrew_how_2008, voci_intergroup_2003, swart_affective_2011}.

Drawing on \citet{schlenker_social_1982}'s social anxiety framework, \citet{plant_antecedents_2003} and \citet{white_improving_2019} suggest that the anxiety individuals feel in social settings is inversely tied to their confidence in creating a favorable impression (i.e., their outcome expectancies). This premise indicates that effectively reducing intergroup anxiety requires a communication environment that fosters outcome expectancies by offering greater autonomy in self-presentation. Computer-mediated communication, such as E-contact, introduces participants with a substantial degree of psychological control over conversational dynamics that is largely absent in direct interactions \cite{amichai-hamburger_contact_2006}. In text-only online interactions, individuals have sufficient time to carefully consider and revise their responses before sending \cite{white_improving_2019}. This structural affordance drives a dual mechanism in which individuals become more assured in their self-presentation skills and experience less uncertainty about appropriate social conduct. Building on longitudinal findings that positive interactions reduce intergroup anxiety by improving outcome expectancies \cite{plant_responses_2004}, \citet{white_improving_2019} demonstrated that E-contact can experimentally trigger this mechanism, thereby enhancing outgroup attitudes.

Alongside intergroup anxiety and outcome expectancies, affective empathy and perspective-taking constitute essential mediating mechanisms in intergroup contact. \citet{davis_empathy_1994} conceptualized empathy as encompassing both the cognitive capacity to adopt others' psychological point of view (i.e., perspective-taking) and the capacity to experience affective reactions to the observed experiences of others (i.e., affective empathy). Engaging in perspective-taking reduces stereotypic biases and ingroup favoritism \cite{galinsky_perspective-taking_2000}, while inducing affective empathy for a member of a stigmatized group can improve attitudes toward the group as a whole \cite{batson_empathy_1997}. Furthermore, \citet{aberson_contact_2007}  established that perspective-taking reduces intergroup anxiety, which in turn predicts more positive outgroup attitudes. They argued that confidence in understanding outgroup perspectives clarifies behavioral norms, thereby reducing intergroup anxiety.

However, conventional perspective-taking interventions carry a structural asymmetry: while dominant group members benefit from adopting outgroup perspectives, non-dominant group members show limited or no improvement under the same approach \cite{bruneau_power_2012}. Evidence from both laboratory and field experiments demonstrates that perspective-giving, compared to perspective-taking, more consistently improves attitudes among non-dominant group members toward the outgroup \cite{bruneau_power_2012, ugarriza_effect_2017}. These findings suggest that effective intergroup interventions should simultaneously facilitate perspective-taking for the dominant group and perspective-giving for the non-dominant group.

Building on these arguments, GroupEnvoy's text-based, synchronous design may afford participants similar psychological control over self-presentation to that in E-contact \cite{amichai-hamburger_contact_2006}, potentially activating outcome expectancies and anxiety-reduction mechanisms. Moreover, its conversational format may naturally promote perspective-taking and affective empathy through sustained engagement with outgroup perspectives. 

E-contact is not intended to replace direct interactions but rather to serve as an important facilitating component within a broader, integrated intervention strategy \cite{white_improving_2015, husnu_elaboration_2010}. While E-contact may be an effective preparatory intervention that can promote future direct contact, reduced intergroup anxiety and improved outcome expectancies do not guarantee that participants will develop stronger intentions to pursue direct interactions, as behavioral intentions serve as the immediate precursor of actual behavior \cite{ajzen_theory_1991}. Drawing on the mediating mechanisms identified in prior literature, we hypothesize that AI-mediated indirect contact via GroupEnvoy will reduce intergroup anxiety while enhancing outcome expectancies, outgroup attitudes, and future contact intentions.

\subsection{Conversational Agents in Group Discussions}

While traditional Computer-Mediated Communication (CMC) operates as a transparent channel, Artificial Intelligence-Mediated Communication (AI-MC) employs intelligent agents to actively modify, augment, or generate messages on behalf of communicators \cite{hancock_ai-mediated_2020, jakesch_ai-mediated_2019}. This capacity to actively shape communicative content has driven HCI research toward the design of agents that facilitate collective deliberation.

Since group discussions often suffer from reasoning failures like groupthink \cite{janis_groupthink_1982} and group polarization \cite{moscovici_group_1969}, HCI researchers deploy conversational agents as neutral mediators, ranging from basic moderators \cite{kim_moderator_2021} to LLM-powered systems that synthesize shared positions across divided groups \cite{tessler_ai_2024, konya_using_2025}. While these approaches reduce social anxiety and facilitator bias, they present outgroup perspectives in a neutralized, aggregated form.  Consequently, they fail to engage the mediating mechanisms, such as affective empathy and perspective-taking, that are essential for reducing prejudice \cite{pettigrew_how_2008}.

Another line of research positions conversational agents not as neutral facilitators but as assertive interlocutors. Drawing on the devil's advocate method, LLM-powered agents can challenge majority opinions \cite{chiang_enhancing_2024} and generate counterarguments to amplify minority opinions \cite{lee_conversational_2025, lee_amplifying_2025}. While these interventions successfully mitigate groupthink and improve intragroup decision quality, they do not address improving intergroup relations through affective engagement.

GroupEnvoy advances this line of research by applying intergroup contact theory. Rather than serving as neutral mediators or intragroup advocates, GroupEnvoy acts as an outgroup representative in discussions among ingroup members. This design exposes ingroup members to authentic outgroup perspectives, thereby directly engaging the mediating mechanisms that prior AI-mediated systems have largely left unaddressed.

\section{Design and Implementation of GroupEnvoy} \label{sec:method}

\subsection{Design Rationales} \label{sec:design_rationales}
To address RQ2, we derived the following design rationales (DR1--DR3) from intergroup contact theory and prior E-contact research. These rationales guided the design of GroupEnvoy, and their fulfillment and limitations are evaluated to derive design implications for AI-mediated intergroup contact in Section~\ref{sec:design_implications}. Table~\ref{tab:design_rationales} summarizes the sub-elements and definitions of each design rationale.

\begin{table*}[t]
\caption{Design rationales and their sub-elements for AI-mediated intergroup contact.}
\label{tab:design_rationales}
\begin{tabular}{p{0.08\textwidth} p{0.16\textwidth} p{0.68\textwidth}}
\toprule
\textbf{DR} & \textbf{Sub-element} & \textbf{Definition} \\
\midrule
\multirow{2}{*}{DR1}
  & Equal Status & The agent is perceived as an equal, proactive participant driven by outgroup values, rather than as a subordinate tool exhibiting sycophantic compliance \cite{pettigrew_intergroup_1998, ma_towards_2024}. \\
  & Cooperation & Both parties collaboratively pursue common goals without role polarization or adversarial dynamics \cite{pettigrew_intergroup_1998, imperato_allport_2021}. \\
\midrule
\multirow{3}{*}{DR2}
  & Typicality & Perceiving the contact partner as a typical outgroup member enables attitude change to generalize to the broader outgroup \cite{hewstone_contact_1986, brown_integrative_2005}. \\
  & Individuation & Self-disclosure and personal information enable person-based evaluations that override stereotypic categorization \cite{miller_personalization_2002, harwood_grandparent-grandchild_2005}. \\
  & Authenticity & The agent's statements are traceable to the authentic perspectives of outgroup members, rather than relying on generative predictions \cite{sharma_towards_2024, chiang_enhancing_2024}. \\
\midrule
\multirow{2}{*}{DR3}
  & Perspective-taking & The interaction promotes participants' cognitive engagement with outgroup viewpoints, reducing intergroup anxiety and stereotypic bias \cite{aberson_contact_2007, galinsky_perspective-taking_2000, davis_empathy_1994}. \\
  & Affective Empathy & The interaction induces emotional resonance with outgroup experiences to improve attitudes toward the broader outgroup \cite{batson_empathy_1997}. \\
\bottomrule
\end{tabular}
\end{table*}

\paragraph{\textbf{DR1. Maintain equal status and cooperation.}} \label{sec:dg1}
AI-mediated intergroup contact requires that the conversational agent be perceived as an equal counterpart rather than a subordinate tool, and that the interaction sustain genuine \textit{cooperation} rather than inflexible or one-sided exchanges. Among Allport's four optimal conditions, \textit{common goals} and \textit{institutional support} should be structurally guaranteed through task design appropriate to the target intergroup context. However, achieving the remaining two conditions, \textit{equal status} \cite{pettigrew_intergroup_1998, wang_reducing_2020} and \textit{cooperation} \cite{pettigrew_intergroup_1998, imperato_allport_2021, amichai-hamburger_contact_2006}, poses distinct challenges when the interlocutor is a conversational agent.

Humans tend to treat AI as a subordinate tool rather than an equal counterpart \cite{zhang_ideal_2021}. This tendency is reinforced by AI sycophancy, in which agents align with users' views rather than offering counterarguments \cite{sharma_towards_2024}. Such compliance casts the agent as a subservient assistant rather than an autonomous interlocutor, thereby directly undermining \textit{equal status}. Furthermore, sustaining meaningful \textit{cooperation} is difficult even when \textit{equal status} is established. During group discussions, AI's inability to track discussion dynamics often leads to inflexible responses and hinders genuine collaboration \cite{zheng_competent_2023}, thereby demonstrating that traditional human-human contact conditions do not seamlessly translate to AI-mediated interactions. We therefore redefine these two conditions for AI-mediated interactions. 
\textit{Equal status} requires treating the agent as an equal counterpart \cite{ma_towards_2024} and recognizing GroupEnvoy as a proactive participant driven by outgroup values rather than demonstrating sycophantic compliance. \textit{Cooperation} involves collaboratively pursuing common goals without role polarization or adversarial dynamics. This directly addresses the inflexibility identified in prior human-AI teaming.

\paragraph{\textbf{DR2. Faithfully convey outgroup perspectives.}} \label{sec:dg2}
For the effects of AI-mediated contact to generalize beyond the immediate interaction, participants should perceive the agent as a \textit{typical yet individuated} outgroup member whose statements faithfully reflect the authentic perspectives of outgroup members. Contact effects only generalize when group membership remains salient, and the partner is viewed as \textit{typical} \cite{hewstone_contact_1986}, preventing their dismissal as a mere exception \cite{johnston_cognitive_1992, wilder_intergroup_1984}. However, \textit{typicality} alone is insufficient. Without personal knowledge of the contact partner, interactions remain governed by stereotype-based judgments rather than genuine interpersonal engagement. \textit{Individuation} through self-disclosure and personal information enables person-based evaluations that override stereotypic categorization \cite{miller_personalization_2002, harwood_grandparent-grandchild_2005}. Yet, excessive individuation risks reducing group salience and undermining the generalization that \textit{typicality} supports. Optimal contact thus requires simultaneously maintaining high perceived outgroup \textit{typicality} and interpersonal \textit{individuation} \cite{brown_integrative_2005, miller_personalization_2002}.


Online communication offers a unique advantage in resolving the inherent tension between \textit{typicality} and \textit{individuation} \cite{hewstone_contact_1986, miller_personalization_2002}. By removing the overwhelming visual category cues present in direct interactions, online environments prevent immediate stereotyping. Consequently, group membership can be made explicitly salient through deliberate design choices, while a shared interpersonal identity emerges from the interaction itself \cite{amichai-hamburger_contact_2006}.

For this collective framing to be credible, GroupEnvoy's content should reflect the authentic perspectives of outgroup members rather than relying solely on generative predictions. Large Language Models (LLMs) are highly susceptible to \textit{sycophancy}, the tendency to prioritize user beliefs over truthful ones \cite{sharma_towards_2024, malmqvist_sycophancy_2025}. This tendency undermines the intergroup contact mechanism by failing to introduce authentic outgroup perspectives. Furthermore, prior work on AI devil's advocate systems demonstrates that AI-generated counterarguments often lack perceived authenticity \cite{chiang_enhancing_2024, lee_amplifying_2025}. Therefore, it is essential for GroupEnvoy to ensure that its statements are traceable to the authentic perspectives of outgroup members.

\paragraph{\textbf{DR3. Promote perspective-taking and affective empathy toward the outgroup.}} \label{sec:dg3}
AI-mediated contact should actively promote participants' own perspective-taking and affective empathy, rather than merely delivering outgroup information for passive reception. As established in Section \ref{sec:mediating_mechanisms}, perspective-taking promotes confidence in understanding outgroup viewpoints while reducing intergroup anxiety \cite{aberson_contact_2007}, stereotypic biases, and ingroup favoritism \cite{galinsky_perspective-taking_2000}, whereas affective empathy has been shown to improve attitudes toward the broader outgroup \cite{batson_empathy_1997}. However, because perspective-taking reduces prejudice through effortful cognitive engagement rather than passive text exposure \cite{davis_empathy_1994, broockman_durably_2016}, proactively supplying outgroup perspectives risks allowing users to bypass this inferential effort, thereby substituting the agent's output for their own perspective-taking. GroupEnvoy's conversational format is therefore designed to facilitate sustained engagement with outgroup perspectives, prompting participants to actively reflect on outgroup viewpoints through dialogue rather than passively receiving information.

\subsection{System Implementation} \label{sec:implementation}

GroupEnvoy's responses were grounded in empirical data collected from a single outgroup session conducted prior to the main experiment (Section~\ref{sec:procedure}). In this session, three outgroup participants engaged in a structured group discussion, from which the following three types of data were extracted and compiled into GroupEnvoy's system prompt. The complete system prompt is provided in the supplementary material.
\begin{itemize}
    \item \textit{Persona data:} Self-introductions and personal experiences relevant to the discussion topic, composed by each participant (approximately 200 Chinese characters per participant).
    \item \textit{Outgroup action plan:} The final collective action plan developed by the group, along with the individual drafts that informed its development.
    \item \textit{Discussion transcripts:} Anonymized full transcripts of the outgroup discussion.
\end{itemize}
These data sources were provided as contextual grounding for GroupEnvoy, which was implemented using \textit{gemini-3-pro-preview} \cite{gemini_team_gemini_2025} via a custom-built web application.

To promote \textit{equal status} under DR1, GroupEnvoy was designed to represent the same number of outgroup members as ingroup participants, thereby functioning as an equal participant rather than a subordinate assistant. Proactivity for \textit{equal status} was maintained through two mechanisms: maintaining the outgroup perspectives and prohibiting affirmative responses when ingroup proposals contradicted the supporting data \cite{sharma_towards_2024}; and posing at most one question per turn from the outgroup viewpoint to prevent unidirectional discussion. To promote \textit{cooperation}, GroupEnvoy was instructed to prioritize the shared \textit{common goal} and maintain a collaborative stance throughout the interaction.

To satisfy DR2, GroupEnvoy was designed to distinguish between information directly supported by the provided data and inferences derived from it. When addressing topics outside the provided data, GroupEnvoy prefaced its statements with a standardized disclaimer (e.g., \textit{``This point was not explicitly discussed among the outgroup members, but based on their backgrounds, I infer that \ldots''}). Outgroup perspectives were further presented in two modes: a \textit{collective mode} (e.g., \textit{``Among the members I represent \ldots''}) for views shared across participants, and an \textit{individual mode} (e.g., \textit{``One member who \ldots''}) for participant-specific perspectives. This design simultaneously supported \textit{typicality} and \textit{individuation}.

To promote \textit{affective empathy} under DR3, GroupEnvoy was instructed to incorporate specific persona references when elaborating on individual backgrounds (e.g., citing a member's self-introduction or unique personal experience), anchoring outgroup perspectives in individuals rather than an abstract collective. The agent was further programmed to maintain a collaborative and empathetic conversational tone while avoiding sycophantic endorsement of ingroup positions that contradicted outgroup perspectives \cite{sharma_towards_2024}. To promote \textit{perspective-taking}, GroupEnvoy was instructed to pose at most one question per turn from the outgroup viewpoint (e.g., \textit{``Is there anything about their hobbies or experiences that caught your attention, or that you’d like to ask them about?''}).

\begin{figure}[!htb]
  \centering
  \includegraphics[width=\columnwidth]{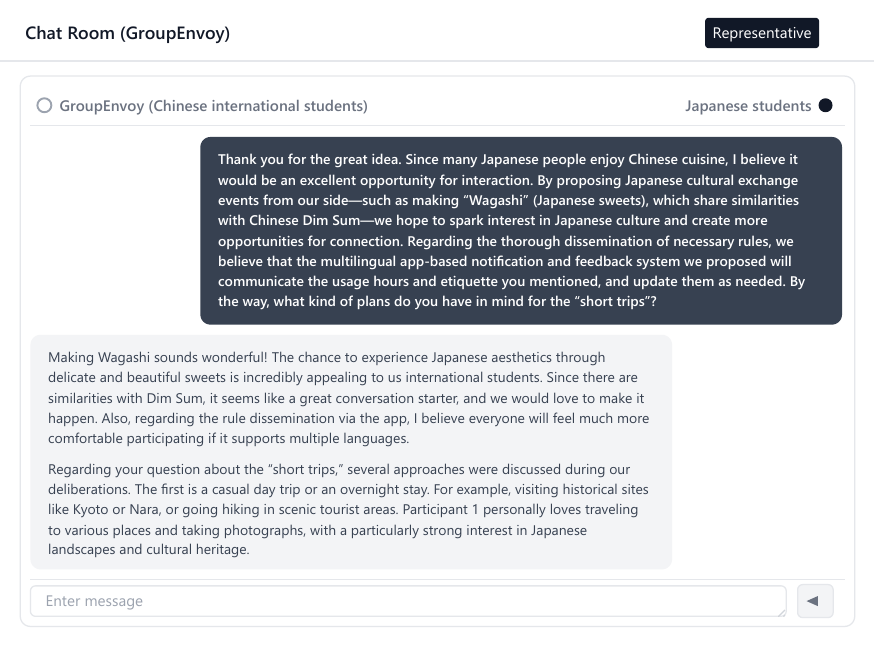}
  \caption{Screenshot of the GroupEnvoy chat interface, translated from Japanese to English. Messages from GroupEnvoy appear on the left, and those from ingroup participants appear on the right. In this example, GroupEnvoy responds to a proposal about cultural exchange events by elaborating on shared interests and referencing individual outgroup members' perspectives.}
  \Description{Screenshot of the GroupEnvoy chat interface. The interface displays a text-based chat room with two columns: messages from GroupEnvoy (representing Chinese international students) appear on the left side with a light background, and messages from Japanese students appear on the right side with a dark background. The header reads "Chat Room (GroupEnvoy)" with a "Representative" badge and "Logout" button. In the conversation shown, the Japanese students propose a cultural exchange event involving Japanese sweets (Wagashi), and GroupEnvoy responds enthusiastically, elaborating on the similarities between Wagashi and Chinese Dim Sum, and sharing individual outgroup members' preferences for short trips to historical sites. A text input field with a send button is located at the bottom of the interface.}
  \label{fig:chat_interface}
\end{figure}

\section{Experiment} \label{sec:experiment}

\subsection{Study Design} \label{sec:study_design}

This study employed a between-subjects design to examine whether interactive dialogue with an AI agent representing outgroup perspectives affects ingroup participants' psychological responses differently from passive exposure to the same content. Figure~\ref{fig:teaser} provides an overview of this design. In the experimental condition, ingroup participants engaged in a collaborative planning task while interacting with GroupEnvoy through a text-based chat interface. In the control condition, participants completed the same collaborative task while reviewing the same outgroup information as a structured, static document. While both conditions introduced identical outgroup source content collected from outgroup members, they differed in whether participants accessed this content through interactive dialogue with GroupEnvoy or as passive written material, thereby isolating \textit{interactivity} as the key manipulated variable under investigation \cite{pereira_da_costa_does_2024}. All sessions followed a three-phase structure: ingroup-only discussion (Phase~1), intergroup contact according to assigned condition (Phase~2), and final ingroup-only discussion (Phase~3). As the host society's linguistic and demographic majority, Japanese students constituted the outgroup (i.e., the dominant group) in this experiment, consistent with evidence that members of the dominant group benefit from perspective-taking (Section~\ref{sec:mediating_mechanisms}).

\subsection{Materials} \label{sec:materials}

\subsubsection {Task and Scenario}

Both conditions shared the same collaborative task and scenario. Participants assumed the role of members of a ``Coexistence Support Room'', an organization established within a fictional Japanese university international dormitory. The organization comprised equal numbers of Japanese students and Chinese international students, and its purpose was to address potential conflicts arising from differences in lifestyle, culture, and language between the two groups. Participants were tasked with collaboratively developing an action plan that reflected both groups' perspectives and was mutually acceptable. Each action plan consisted of two components: a \textit{measures} section (280--320 Japanese characters) and a \textit{rationale} section (unconstrained length). The shared objective structurally instantiated Allport's \textit{common goals} condition, while the organizational framing of the ``Coexistence Support Room'' served as a form of \textit{institutional support} for the contact situation \cite{allport_nature_1954, pettigrew_intergroup_1998}.

\subsubsection {Outgroup Data}

Outgroup data were collected from outgroup members in a single session conducted prior to the main experiment (Section~\ref{sec:outgroup_session}). Three types of data (i.e., \textit{persona data}, \textit{outgroup action plan}, and \textit{discussion transcripts}) were retained as described in Section \ref{sec:implementation}. All data were originally recorded in Chinese and used across all experimental sessions.

\subsubsection {Condition-Specific Materials}

In the experimental condition, the outgroup data were retained in their original Chinese and incorporated into GroupEnvoy's system prompt, as described in Section~\ref{sec:implementation}. GroupEnvoy processed these inputs and generated responses in Japanese. In the control condition, participants received the same outgroup information as a structured document during Phase~2. All original materials were translated into Japanese by the first author, a native speaker of both languages. The document was organized into four sections: persona information, individual and collective action plans, discussion transcripts, and full discussion transcripts. Since the full discussion transcripts were lengthy and difficult to review within the session time, the summarized version was produced by adapting the prompting strategy of \citet{kim_nexussum_2025} with \textit{gemini-3-pro-preview} \cite{gemini_team_gemini_2025}.

\subsection{Participants} \label{sec:participants}

We recruited 121 Japanese university and graduate students via social networking services. Applicants were screened to exclude those who (1) had prior close friendships with Chinese international students, (2) had lived or studied in China, or (3) had formal experience studying Chinese, leaving 60 eligible applicants. These criteria were designed to exclude participants likely to have more positive baseline outgroup attitudes while avoiding restrictive screening, given the high prevalence of Chinese acquaintances in Japan. 18 participants were randomly assigned to six groups of three and to either the experimental or control condition (three groups per condition).\footnote{One participant in Group~6 withdrew shortly before the session, so this group completed the experiment with two members.} The final sample comprised 17 Japanese students: 9 in the experimental condition ($M = 23.67$, $SD = 3.54$) and 8 in the control condition ($M = 24.62$, $SD = 4.07$). The outgroup comprised 3 Chinese international students enrolled at Japanese universities ($M = 20.67$, $SD = 2.89$) who participated in a single outgroup session to generate the grounding data for GroupEnvoy. All participants provided written informed consent before participation and were compensated with a 3700 JPY gift card. This study was reviewed and approved by the Ethical Review Committee for Experimental Research involving Human Subjects at the University of Tokyo.

\begin{table}[t]
  \caption{Demographics of all participants. Contact: whether the participant reported having Chinese acquaintances other than international students.}
  \label{tab:participants}
  \centering
  \begin{tabular}{lllll}
    \toprule
    Group & ID & Age & Gender & Contact \\
    \midrule
    \multirow{9}{*}{Experimental}
      & EP1 & 24 & F & No \\
      & EP2 & 27 & M & No \\
      & EP3 & 21 & F & No \\
      & EP4 & 22 & M & No \\
      & EP5 & 25 & M & No \\
      & EP6 & 31 & M & No \\
      & EP7 & 22 & F & No \\
      & EP8 & 21 & F & No \\
      & EP9 & 20 & F & Yes \\
    \midrule
    \multirow{8}{*}{Control}
      & CP1 & 26 & M & No \\
      & CP2 & 20 & F & No \\
      & CP3 & 21 & F & No \\
      & CP4 & 30 & F & No \\
      & CP5 & 31 & F & Yes \\
      & CP6 & 24 & F & No \\
      & CP7 & 22 & M & No \\
      & CP8 & 23 & M & No \\
    \midrule
    \multirow{3}{*}{Outgroup}
      & P1 & 19 & F & -- \\
      & P2 & 24 & M & -- \\
      & P3 & 19 & M & -- \\
    \bottomrule
  \end{tabular}
\end{table}

\begin{figure*}[t]
  \centering
  \includegraphics[width=\textwidth]{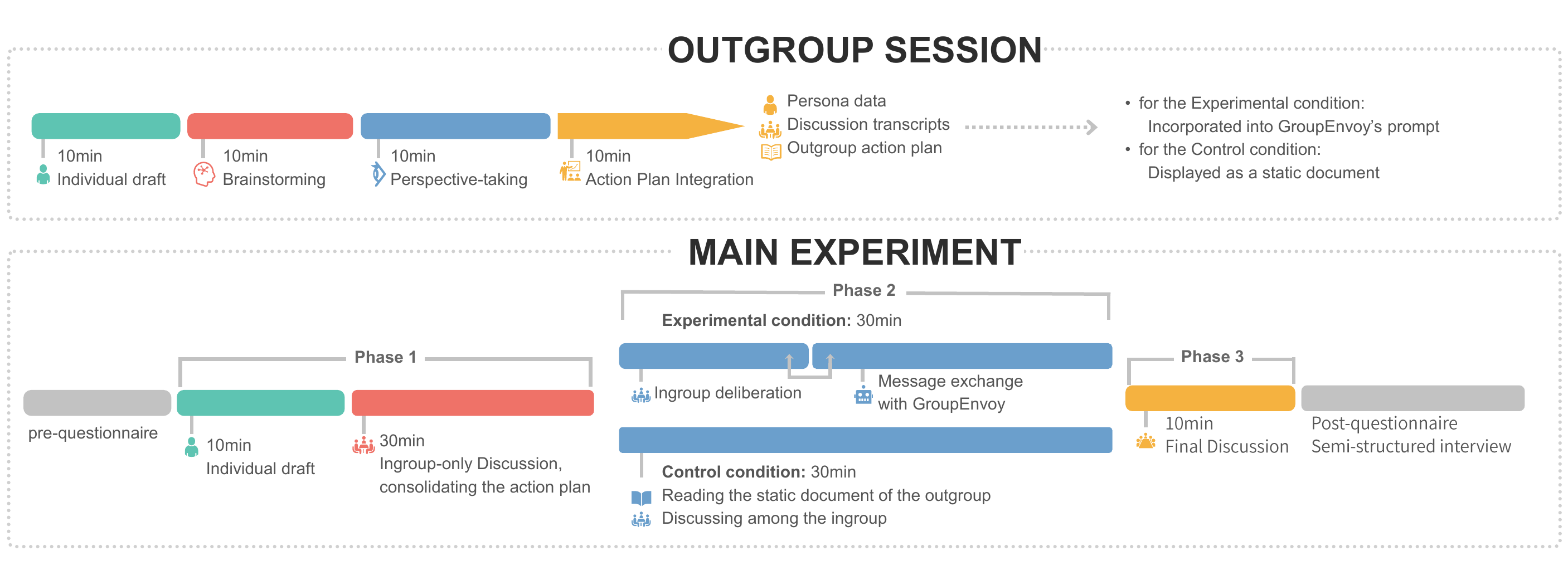}
  \caption{Overview of the experimental procedure.}
  \Description{A horizontal flowchart of the study procedure with three phases. Phase 1, the Outgroup Session, shows three outgroup participants drafting individual plans and discussing through brainstorming, perspective-taking, and consolidation, producing persona data, an outgroup action plan, and discussion transcripts that feed into GroupEnvoy's prompt. Phase 2, the Main Experiment, shows ingroup participants moving through numbered steps: a pre-experiment questionnaire and individual draft, an ingroup-only discussion, and an intergroup contact step that branches into an experimental path (message exchange with GroupEnvoy) and a control path (reading a static document of the outgroup data), then a final ingroup discussion. Phase 3 shows a post-experiment questionnaire and a semi-structured interview.}
  \label{fig:procedure}
\end{figure*}

\subsection{Procedure} \label{sec:procedure}
The study was conducted entirely online via Zoom\footnote{\url{https://zoom.us/}} and proceeded in two phases: an outgroup session and main sessions with ingroup participants. Figure~\ref{fig:procedure} presents an overview of the full procedure.

\subsubsection{Outgroup Session} \label{sec:outgroup_session}
The three outgroup participants convened for a structured discussion session. After individually drafting preliminary action plans (10min), one participant was designated as facilitator. The discussion proceeded through three stages: open brainstorming (20 min), a perspective-taking activity in which they were asked to adopt Japanese students' perspectives (10 min), and consolidation of a collective action plan (10 min). The perspective-taking activity ensured that the grounding data reflected not only outgroup perspectives but also the outgroup engagement with ingroup perspectives. Three types of data were extracted from this session (persona information, action plan, and discussion transcripts) and compiled into GroupEnvoy's system prompt (Section~\ref{sec:implementation}). The outgroup action plan used the same two-part structure as the ingroup's, with the \textit{measures} section set at 140--160 characters.\footnote{Chinese characters convey substantially more information per character than Japanese, so fewer characters were required to express comparable content.}

\subsubsection{Main Experiment} \label{sec:main_experiment}
Each ingroup session followed four steps, preceded by individual pre-experiment questionnaires and the drafting of preliminary action plans (10 min). One member per group was designated as the facilitator, responsible for managing the discussion flow and, in the experimental condition, submitting messages to GroupEnvoy on behalf of the group's collective deliberation.

\begin{enumerate}
  \item Ingroup-Only Discussion (30 min). The group engaged in open-ended brainstorming and developed a shared preliminary action plan without access to outgroup information.
  \item Intergroup Contact (30 min). After a brief introduction of the condition-specific interface, participants engaged with outgroup perspectives for 30 minutes. In the experimental condition, the facilitator logged into the GroupEnvoy web application. Participants began with self-introductions and brief small talk directed to GroupEnvoy, then the facilitator shared their preliminary action plan. After GroupEnvoy introduced the outgroup action plan, participants engaged in iterative dialogue: deliberating among themselves in light of GroupEnvoy's responses, then having the facilitator compose and send a message to GroupEnvoy. All responses were displayed in real time on participants' screens. In the control condition, participants received a structured document containing the same outgroup data and spent the session reading and discussing its contents.
  \item Final Discussion (10 min). Participants revised and finalized their action plan in an intragroup discussion.
  \item Post-Experiment. To evaluate perceived psychological change and condition-specific experiences, participants completed a post-experiment questionnaire, followed by a semi-structured group interview.
\end{enumerate}

\subsubsection{Outgroup Review} \label{sec:outgroup_review}
To assess the authenticity of GroupEnvoy's responses (DR2), we conducted a post-hoc review with outgroup members. Two of the three outgroup participants (P1 and P2) consented to this additional procedure. The two reviewers independently evaluated each of GroupEnvoy's 30 responses from the experimental sessions, classifying each as: (1) \textit{explicitly stated} in their original discussion, (2) \textit{aligned} with their views though not explicitly discussed, or (3) \textit{neither}.

\subsection{Measurements} \label{sec:measurements}

All quantitative measures were rated on a seven-point Likert scale (1 = \textit{strongly disagree}, 7 = \textit{strongly agree}) unless otherwise noted and were administered before and after the experiment.

To evaluate RQ1, we measured the following four measures.

\begin{itemize}
    \item \textbf{Intergroup anxiety} was assessed using a six-item bipolar scale adapted from \citet{stephan_intergroup_1985}. Responses were anchored by bipolar adjective pairs: \textit{relaxed--nervous}, \textit{pleased--worried}, \textit{not scared--scared}, \textit{at ease--awkward}, \textit{open--defensive}, and \textit{confident--unconfident} (higher scores indicate greater anxiety).
    \item \textbf{Outcome expectancies} were assessed with an eleven-item scale adapted from \citet{plant_antecedents_2003}, as used in \citet{white_improving_2019} (e.g., ``I am confident that I can respond in a nonprejudiced manner toward Chinese international students'').
    \item \textbf{Future contact intentions} toward Chinese international students and Chinese people in general were assessed using a 7-item scale adapted from \citet{husnu_elaboration_2010}.
    \item \textbf{Outgroup attitudes} toward both the Chinese people in general and Chinese international students were assessed using a four-item semantic differential scale adapted from \citet{wright_extended_1997}. Responses were anchored by bipolar adjective pairs: \textit{negative--positive}, \textit{hostile--friendly}, \textit{suspicious--trusting}, and \textit{contemptuous--respectful} (higher scores indicate more positive attitudes).

\end{itemize}

To assess the degree to which GroupEnvoy facilitated the mediating mechanisms targeted by DR3 (Section~\ref{sec:dg3}), we additionally measured the following two measures.

\begin{itemize}
    \item \textbf{Perspective-taking} was assessed with a six-item intergroup understanding scale adapted from \citet{stephan_walter_g_survey_1999}, as used in \citet{aberson_contact_2007} (e.g., ``I believe that I have a good understanding of how Chinese international students view the world'').
    \item \textbf{Affective empathy} toward Chinese people was assessed using a three-item scale adapted from \citet{swart_affective_2011} (e.g., \textit{''When a Chinese friend of mine is sad, I feel sad too''}).
\end{itemize}

Following the experiment, semi-structured group interviews were conducted to explore participants' psychological changes (RQ1). Each question was posed in sequence, with participants invited to respond in turn. Participants in the experimental condition additionally evaluated GroupEnvoy against its design rationales (RQ2). All sessions were audio-recorded with participants' consent and subsequently transcribed.

\subsection{Analysis} \label{sec:analysis}

For each psychological measure, we conducted a 2 (Group: Experimental vs. Control) $\times$ 2 (Time: pre vs. post) mixed ANOVA with Group as a between-subjects factor and Time as a within-subjects factor. Mixed ANOVA was selected to assess Group × Time interaction effects, enabling direct comparison of pre-to-post change between conditions. Partial eta squared ($\eta_p^2$) was calculated to estimate the effect size. We also applied Template Analysis \cite{ciesielska_qualitative_2018,brooks_utility_2015}, a codebook approach to thematic analysis \cite{braun_toward_2023} that supports theory-derived a priori themes, to the post-experiment semi-structured group interview transcripts from both conditions. The initial template was built from a priori themes (i.e., the measured psychological constructs for RQ1, and the design rationales DR1--DR3 for RQ2). This initial template was iteratively refined on the experimental condition data, yielding the experimental condition template with 13 top-level themes and 60 lower-level codes. To enable a common-basis comparison, the control condition template was derived from the experimental condition template by removing themes specific to the experimental condition and adding those unique to the control condition, resulting in 12 top-level themes and 52 lower-level codes. The first author segmented and coded all transcripts, and the second author independently coded those of one group from the experimental condition. Inter-coder reliability was then assessed: Cohen's $\kappa=.61$ at the lowest code level, and $\kappa=.84$ at the top-level theme level. All disagreements were subsequently resolved through discussion between the two coders. Additionally, two outgroup members independently evaluated the authenticity of GroupEnvoy's response (DR2) following the procedure described in Section~\ref{sec:outgroup_review}.


\section{Quantitative Results} \label{sec:quant_results}

\begin{figure*}[t]
  \centering
  \includegraphics[width=\textwidth]{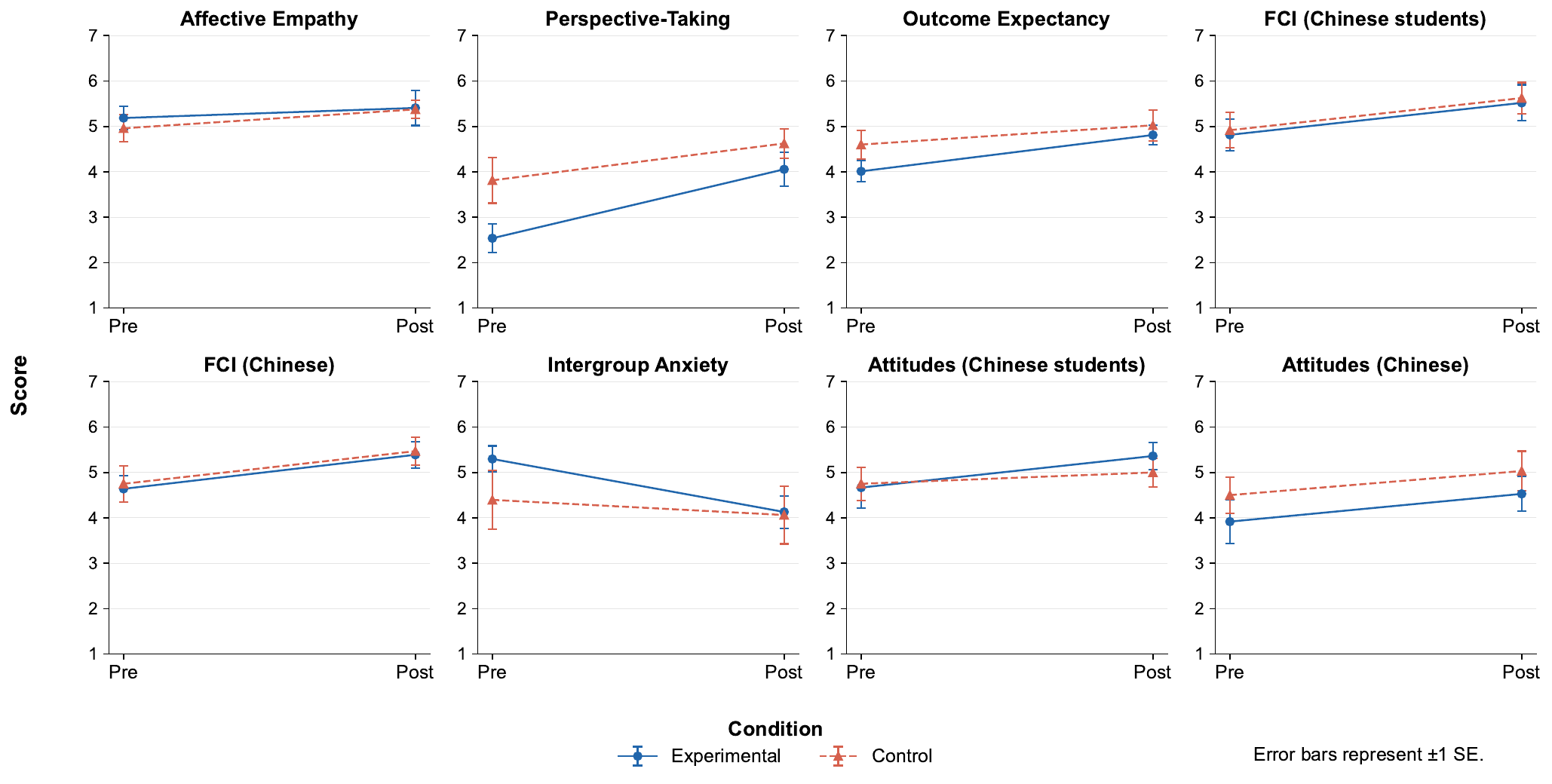}
  \caption{Mean scores (with $\pm 1$ SE error bars) for each psychological measure at pre- and post-test, by condition. All measures used a 7-point Likert scale. FCI: Future Contact Intentions. Attitudes: Outgroup attitudes.}
  \Description{A 2×4 grid of line plots showing pre-to-post changes in eight psychological measures by condition (Experimental in blue solid lines, Control in red dashed lines), each on a 1–7 Likert scale with ±1 SE error bars. Top row, left to right: Affective Empathy (Experimental: 5.19→5.41, Control: 4.96→5.38), Perspective-Taking (Experimental: 2.54→4.06, Control: 3.81→4.63), Outcome Expectancy (Experimental: 4.01→4.81, Control: 4.60→5.03), FCI for Chinese international students (Experimental: 4.81→5.52, Control: 4.92→5.62). Bottom row: FCI for Chinese (Experimental: 4.64→5.39, Control: 4.75→5.47), Intergroup Anxiety (Experimental: 5.30→4.13, Control: 4.40→4.06), Attitudes toward Chinese international students (Experimental: 4.67→5.36, Control: 4.75→5.00), Attitudes toward Chinese (Experimental: 3.92→4.53, Control: 4.50→5.03). The most visually prominent interaction patterns appear for Perspective-Taking (Experimental Δ=+1.52 vs. Control Δ=+0.82) and Intergroup Anxiety (Experimental Δ=−1.17 vs. Control Δ=−0.34), where the experimental condition shows larger pre-to-post changes than the control.}
  \label{fig:interaction}
\end{figure*}

We conducted a 2 (Group: Experimental vs.\ Control) $\times$ 2 (Time: Pre vs.\ Post) mixed ANOVA for each of the eight psychological measures, with the significance level set at $\alpha = .05$. Figure~\ref{fig:interaction} displays the pre-to-post changes in each measure by condition, and Table~\ref{tab:anova} presents the full descriptive statistics and ANOVA results.

\begin{table*}[t]
  \caption{%
    Descriptive statistics ($M$, $SD$) and mixed ANOVA results.
    $\eta_p^2$ : partial eta squared.
    Exp: Experimental condition. Ctrl: Control condition.
    $^\dagger p < .10$;\ $^*p < .05$;\ $^{**}p < .01$;\ $^{***}p < .001$.
  }
  \label{tab:anova}
  \resizebox{\textwidth}{!}{%
  \begin{tabular}{l
      cc cc cc cc
      ccc ccc ccc}
    \toprule
    & \multicolumn{2}{c}{\textbf{Exp Pre}}
    & \multicolumn{2}{c}{\textbf{Exp Post}}
    & \multicolumn{2}{c}{\textbf{Ctrl Pre}}
    & \multicolumn{2}{c}{\textbf{Ctrl Post}}
    & \multicolumn{3}{c}{\textbf{Group}}
    & \multicolumn{3}{c}{\textbf{Time}}
    & \multicolumn{3}{c}{\textbf{Group $\times$ Time}} \\
    \cmidrule(lr){2-3}\cmidrule(lr){4-5}
    \cmidrule(lr){6-7}\cmidrule(lr){8-9}
    \cmidrule(lr){10-12}\cmidrule(lr){13-15}\cmidrule(lr){16-18}
    \textbf{Measure}
      & $M$ & $SD$ & $M$ & $SD$ & $M$ & $SD$ & $M$ & $SD$
      & $F(1,\text{15})$ & $p$ & $\eta_p^2$
      & $F(1,\text{15})$ & $p$ & $\eta_p^2$
      & $F(1,\text{15})$ & $p$ & $\eta_p^2$ \\
    \midrule
    Affective Empathy
      & 5.19 & 0.78 & 5.41 & 1.16 & 4.96 & 0.84 & 5.38 & 0.55
      & 0.11 & .743 & .007
      & 3.56 & $.079^\dagger$ & .192
      & 0.33 & .574 & .022 \\
    Perspective-taking
      & 2.54 & 0.94 & 4.06 & 1.13 & 3.81 & 1.42 & 4.63 & 0.92
      & 3.69 & $.074^\dagger$ & .198
      & 21.14 & $<$.001$^{***}$ & .585
      & 1.94 & .184 & .114 \\
    Outcome Expectancies
      & 4.01 & 0.70 & 4.81 & 0.63 & 4.60 & 0.89 & 5.03 & 0.96
      & 1.30 & .272 & .080
      & 14.15 & $.002^{**}$ & .486
      & 1.33 & .267 & .081 \\
    FCI (Chinese students)
      & 4.81 & 1.04 & 5.52 & 1.19 & 4.92 & 1.11 & 5.62 & 0.97
      & 0.05 & .830 & .003
      & 9.85 & $.007^{**}$ & .396
      & 0.00 & .992 & .000 \\
    FCI (Chinese)
      & 4.64 & 0.88 & 5.39 & 0.87 & 4.75 & 1.13 & 5.47 & 0.87
      & 0.06 & .811 & .004
      & 10.02 & $.006^{**}$ & .400
      & 0.00 & .947 & .000 \\
    Intergroup Anxiety
      & 5.30 & 0.87 & 4.13 & 1.07 & 4.40 & 1.83 & 4.06 & 1.80
      & 0.56 & .467 & .036
      & 9.09 & $.009^{**}$ & .377
      & 2.81 & .115 & .158 \\
    Attitudes (Chinese students)
      & 4.67 & 1.35 & 5.36 & 0.90 & 4.75 & 1.03 & 5.00 & 0.90
      & 0.08 & .777 & .005
      & 6.23 & $.025^{*}$ & .293
      & 1.38 & .259 & .084 \\
    Attitudes (Chinese)
      & 3.92 & 1.46 & 4.53 & 1.15 & 4.50 & 1.12 & 5.03 & 1.24
      & 0.87 & .365 & .055
      & 10.28 & $.006^{**}$ & .407
      & 0.05 & .826 & .003 \\
    \bottomrule
  \end{tabular}}
\end{table*}

No measure reached statistical significance for the Group~$\times$~Time interaction. Nevertheless, two measures exhibited notable directional trends consistent with RQ1. Intergroup anxiety showed the largest interaction effect size ($\eta_p^2 = .158$), with the experimental group showing a descriptively greater decrease ($\Delta = -1.17$) than the control group ($\Delta = -0.34$). Perspective-taking also showed a medium interaction effect size ($\eta_p^2 = .114$), with a descriptively larger increase in the experimental group ($\Delta = +1.52$) than in the control group ($\Delta = +0.82$). These effect sizes fall in the medium-to-large range~\cite{cohen_statistical_1988}, suggesting that the lack of statistical significance may reflect insufficient statistical power due to the small sample size rather than the absence of an effect.

A significant main effect of Time was observed for seven of the eight measures: perspective-taking, outcome expectancies, future contact intentions (both targets), intergroup anxiety, and outgroup attitudes (both targets). Affective empathy did not reach significance but showed a marginal trend in the same direction. Both conditions showed consistent improvements, suggesting that exposure to outgroup perspectives contributed to positive psychological change regardless of modality. No significant main effect of Group was observed for any measure. Furthermore, pre-test group differences on certain measures, most notably perspective-taking, may introduce ceiling effects that constrain the rate of change.

\begin{figure*}[t]
  \centering
  \includegraphics[width=\textwidth]{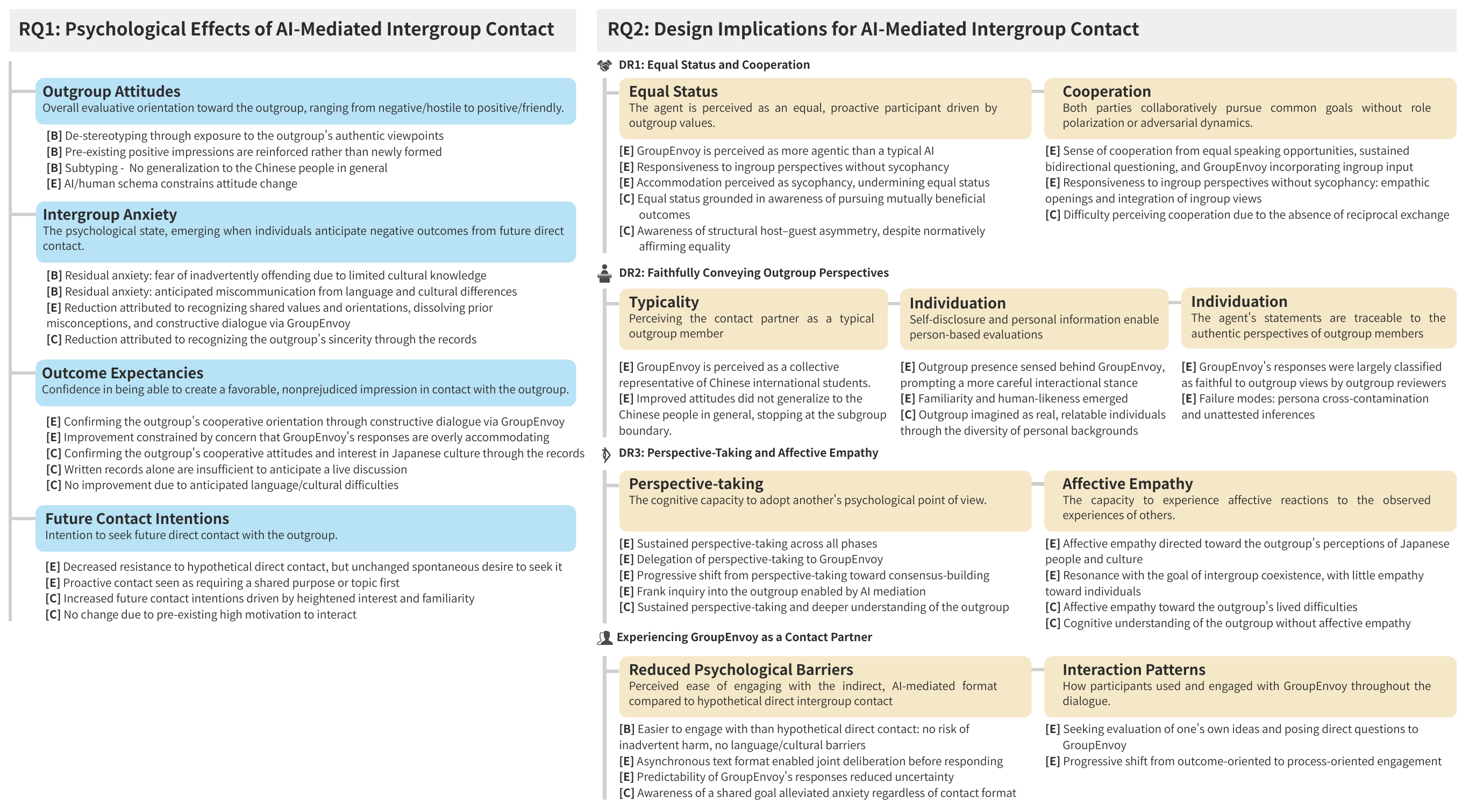}
  \caption{This figure maps the codes identified through template analysis of the semi-structured interviews onto the two research questions. The left column categorizes codes under the four measured psychological constructs (RQ1). The right column organizes codes around the three design rationales and a category that captures participants' experiences with GroupEnvoy (RQ2). Each code carries a tag indicating the condition in which it was identified: [B] for both conditions, [E] for the experimental condition only, and [C] for the control condition only.}
  \label{fig:qualitative_overview}
  \Description{A diagram in two columns summarizing the study's qualitative results. The left column, labeled RQ1, contains four sub-headings for the measured psychological constructs (outgroup attitudes, intergroup anxiety, outcome expectancies, and future contact intentions), each followed by a bulleted list of codes describing the mechanisms of psychological change. The right column, labeled RQ2, contains three design rationales as headings (DR1: equal status and cooperation; DR2: faithfully conveying outgroup perspectives; DR3: perspective-taking and affective empathy) and a fourth heading on experiencing GroupEnvoy as a contact partner; each heading is divided into sub-themes with bulleted lists of codes. Each code is annotated as specific to the Experimental condition, specific to the Control condition, or shared across both conditions.}
\end{figure*}

\section{Qualitative Results} \label{sec:qualitative_results}

\subsection{Experiencing GroupEnvoy as a Contact Partner}

\subsubsection{Reduced Psychological Barriers Compared to Direct Contact}

All participants in the experimental and control conditions reported that the indirect contact format they experienced was easier to engage with than hypothetical direct intergroup contact formats. Participants in both conditions cited two shared sources of ease: engaging with the outgroup without risking inadvertent harm to them, and without facing language and cultural barriers.
\begin{quote}
    \textit{``Since GroupEnvoy is not a human being with feelings, even if I said something inappropriate, the other party would not be hurt.''} \punit{EP9}
\end{quote}

Beyond these shared sources, participants in the experimental condition identified additional factors. EP2 described how the asynchronous text format enabled the three ingroup members to deliberate together before submitting their responses, making the discussion feel more manageable. EP9 added that the predictability of GroupEnvoy's responses reduced uncertainty about how the interaction would unfold. 

In the control condition, some participants \punit{CP5, CP7, CP8} reported no anxiety in either format, noting that the awareness of a shared goal alleviated it. In contrast, CP1 reported experiencing greater anxiety due to the absence of direct exchange.

Taken together, these findings generally align with prior work demonstrating that indirect and media-mediated contact tends to be less anxiety-provoking than direct intergroup contact \cite{amichai-hamburger_contact_2006, white_beyond_2021}.

\subsubsection{Perceived Responsiveness to Ingroup Perspectives Without Sycophancy}

Participants in the experimental condition \punit{EP1, EP2, EP3, EP5, EP7, EP8, EP9} perceived GroupEnvoy as responsive to ingroup values without being sychophantic, attributing this to both its empathic and appreciative openings, and its integration of ingroup perspectives into arguments.
\begin{quote}
    \textit{``The responses seemed to include ideas from the Chinese international students' side, such as holding parties and events, while also aligning somewhat with ours.''} \punit{EP1}
\end{quote}

This responsiveness was generally well-received, with participants \punit{EP8, EP9} reporting a sense of warmth and approachability. However, some participants \punit{EP4, EP6} perceived sycophancy in GroupEnvoy's responses. EP4 expressed concern that accommodating the ingroup would undermine the value of intercultural dialogue.
\begin{quote}
    \textit{``This is a conversation between people from different cultural backgrounds. If the AI accommodates us by ignoring those differences, it defeats the purpose.''} \punit{EP4}
\end{quote}

\subsubsection{Interaction Patterns with GroupEnvoy}

Across all three groups of the experimental condition, participants sought evaluation of their own ideas \punit{EP1, EP2, EP3, EP6, EP7} and posed direct questions to GroupEnvoy \punit{EP1, EP5, EP8}. EP5 noted a shift from evaluating outgroup ideas to exploring the cognitive and emotional processes behind those ideas as the dialogue progressed. Furthermore, participants' engagement progressively shifted from outcome-oriented to process-oriented \punit{EP2, EP3, EP4, EP6, EP7, EP8, EP9}. Instead of using GroupEnvoy for specific outcomes, participants viewed the interaction as an end in itself, allowing them to engage in authentic collaboration and problem-solving.

In two groups of the experimental condition, GroupEnvoy's role in consolidating summaries progressively diminished the ingroup's own sense of initiative in the dialogue \punit{EP2, EP3, EP4, EP5}.

\begin{quote}
    \textit{``After a while, the AI took the initiative to move the discussion forward. From that point on, I left the task of summarizing our opinions to the AI.''} \punit{EP2}
\end{quote}




\subsubsection{Perceived Human-likeness in GroupEnvoy's Responses}

Some participants \punit{EP7, EP8} perceived human-like qualities in GroupEnvoy's responses and viewed the agent favorably. EP8 attributed this quality to empathetic expressions and tangential remarks. In contrast, EP9 perceived a formulaic AI-like pattern in the fixed opening phrase \textit{``Thank you very much''} that appeared at the start of every response.
\begin{quote}
    \textit{``I could tell it was an AI, but the way it added empathetic responses and small bits of information unrelated to the main topic made it feel quite human. I found that warmth appealing.''} \punit{EP8}
\end{quote}

\subsection{Perceived Conditions of Intergroup Contact}

\subsubsection{Perceived Equal Status}

Participants in both conditions generally maintained a perception of \textit{equal status} during the dialogue. In the experimental condition, this perception stemmed from viewing GroupEnvoy as an agentic participant. Some participants \punit{EP1, EP2, EP3, EP6, EP7} described it as more agentic than a typical AI, while others recognized it as having agency equivalent to that of a human. This perceived agency was reflected in specific features of the dialogue, such as a consensus-oriented dialogue \punit{EP2} and mutual respect between the parties \punit{EP3}.
\begin{quote}
    \textit{``Since GroupEnvoy highly respected our proposals and the final action plan was heavily based on our suggestions, I perceived its agency as moderate, yet still greater than typical generative AI.''} \punit{EP3}

    \textit{``In terms of agency, there was no noticeable difference compared to a human participant.''} \punit{EP5}
\end{quote}

However, EP8 perceived GroupEnvoy's response style as accommodating to the ingroup, thereby undermining \textit{equal status}.
\begin{quote}
    \textit{``Rather than perceiving equal status, I felt GroupEnvoy was being somewhat considerate toward us and accommodating us.''} \punit{EP8}
\end{quote}

Participants in the control condition attributed their sense of \textit{equal status} to an awareness of pursuing mutually beneficial outcomes \punit{CP1, CP8}. However, some participants \punit{CP4, CP5, CP7} reported an awareness of structural asymmetry rooted in the host--guest relationship involving Chinese international students in Japan, even as they normatively affirmed that both parties should participate equally.
\begin{quote}
    \textit{``I do think we should be equal, but I faintly held a sense that the Japanese were the inviting side and the Chinese international students were the visiting side.''} \punit{CP4}
\end{quote}

\subsubsection{Perceived Cooperation}

All participants in the experimental condition perceived a sense of \textit{cooperation} with GroupEnvoy during dialogue. In the control condition, only CP6 perceived it, while CP2 held an ambivalent view. Participants in the experimental condition attributed this to equal speaking opportunities for both parties \punit{EP2}, sustained bidirectional questioning \punit{EP3}, and GroupEnvoy's responsiveness in incorporating their input into subsequent replies \punit{EP6}.

The absence of reciprocal exchange in the control condition indicates that participants largely made decisions on their own, making it difficult for them to perceive \textit{cooperation}.
\begin{quote}
    \textit{``It seemed like we just took our original framework, referenced the Chinese international students' views, and revised our plan accordingly.''} \punit{CP8}
\end{quote}

\subsubsection{Perceived Outgroup Presence}

Participants in both the experimental \punit{EP4, EP5, EP7, EP8} and control \punit{EP3, EP4, EP5, EP6, EP7, EP8} conditions broadly reported sensing an \textit{outgroup presence} grounded in the personal and cultural backgrounds of the outgroup. In the experimental condition, this perceived presence directly influenced participants' interactional stance.
\begin{quote}
    \textit{``I felt like there was a person behind it more than usual, so I was more careful about how I listened.''} \punit{EP5}
\end{quote}

Despite this, participants in the experimental condition \punit{EP1, EP2, EP3, EP5, EP6} perceived limited individuation in GroupEnvoy, perceiving it more as a \textit{collective representative} than as distinct individuals.
\begin{quote}
    \textit{``The AI was largely summarizing the Chinese international students' views when it spoke, so I did not find myself thinking much about each Chinese student.''} \punit{EP1}
\end{quote}

By contrast, participants in the control condition \punit{CP3, CP4, CP5, CP6, CP8} reported being able to imagine the Chinese international students as real, relatable individuals due to the diversity of personal backgrounds described in the provided information.
\begin{quote}
    \textit{``Since the scenarios were structured on the premise that various people have different experiences, I could perceive them as real individuals.''} \punit{CP8}
\end{quote}

\subsubsection{Familiarity Through Personal Information and Its Limits}

A factor specific to the experimental condition was the emergence of familiarity and perceived human-likeness through personal information presented via GroupEnvoy, including self-introductions, hobbies, and daily life.
\begin{quote}
    \textit{``Their self-introductions made me feel more familiar with them. I realized that they are just like other Japanese college students.''} \punit{EP8}
\end{quote}

However, this familiarity did not deepen into genuine individual-level understanding for some participants \punit{EP1, EP7}, as the AI-mediated format prevented direct dialogue with outgroup members. 
\begin{quote}
    \textit{``Since the other party was an AI summarizing opinions, I'm not really sure if I can say my understanding of them has truly deepened.''} \punit{EP1}
\end{quote}

In the control condition, no participants reported comparable familiarity. Due to time constraints, participants \punit{CP2, CP4, CP6} ignored these documents and prioritized activity proposals.

\subsection{Empathy, Perspective-Taking, and Attitude Change}

\subsubsection{Affective Empathy}

While participants in both conditions reported some form of empathy toward the outgroup, the target differed qualitatively between conditions. Participants in the experimental condition \punit{EP1, EP4, EP5, EP6, EP7} reported affective empathy directed primarily toward Chinese international students' views of Japanese people and culture. EP4 further noted that Chinese international students' impressions of Japanese students closely aligned with Japanese students' self-perceptions. In contrast, EP2 and EP3 resonated with the goal of intergroup coexistence while stopping short of empathy toward individual Chinese international students.
\begin{quote}
    \textit{``Because GroupEnvoy was speaking on behalf of others, I did not find myself feeling much empathy toward any of them as individuals.''} \punit{EP2}
\end{quote}

In the control condition, participants directed their empathy toward the outgroup experience of difficulty in building social connections \punit{CP4, CP8} and problems of dormitory life \punit{CP6}. 

Some participants in both conditions \punit{EP8, EP9, CP1, CP7} reported cognitive understanding of the outgroup situation without reaching affective empathy.
\begin{quote}
    \textit{``Since I’ve never studied abroad, it’s hard to say I can truly empathize with them, but I felt that I could certainly understand their feelings.''} \punit{EP8}
\end{quote}

\subsubsection{Perspective-taking}

While participants in both conditions broadly engaged in perspective-taking, the experimental condition exhibited three patterns: sustained engagement in perspective-taking throughout the discussion, delegating perspective-taking to GroupEnvoy, and a progressive shift from perspective-taking toward consensus-building as the dialogue progressed.

The first pattern was sustained engagement in perspective-taking across all phases \punit{EP7, EP8, EP9}. EP8 described a deepening trajectory, becoming more outgroup-centered by Phase~3 than in Phase~1.

The second pattern was the delegation of perspective-taking effort to GroupEnvoy \punit{EP1, EP2, EP3}. They treated GroupEnvoy as responsible for representing the outgroup perspective and concentrated on building consensus from their position.
\begin{quote}
    \textit{``When it came to taking the perspective of the Chinese international students, I found myself thinking that GroupEnvoy would handle that for me, and so I did not actively put myself in their position.''} \punit{EP1}
\end{quote}

The third pattern was a progressive shift from perspective-taking toward consensus-building \punit{EP4, EP5, EP6}. Although they entered Phase~2 intending to adopt the outgroup perspective, they gradually shifted toward consensus-building as the dialogue advanced.
\begin{quote}
    \textit{``At first, I was trying to put myself in the position of the Chinese international students, but as GroupEnvoy's responses became more predictable, my desire to reach a good consensus grew stronger.''} \punit{EP6}
\end{quote}

Across these patterns, EP3 and EP5 noted that AI-mediated contact allowed them to ask the outgroup directly on matters that would have felt intrusive in direct contact.

Participants in the control condition reported a deeper understanding of the outgroup \punit{CP1, CP3, CP5, CP6, CP7, CP8}, and sustained perspective-taking \punit{CP1, CP2, CP5, CP6, CP7, CP8}.
\begin{quote}
    \textit{``Since I had no prior interaction with Chinese people or Chinese international students, learning about the reality by reviewing the discussion transcripts makes me believe that my understanding has indeed deepened.''} \punit{CP6}
\end{quote}

\subsubsection{De-stereotyping and Attitude Change toward Chinese International Students} \label{sec:de_stereotyping}

Participants varied in the degree of attitude change: some reported a positive change \punit{EP4, EP8, CP1, CP3, CP5, CP8}, while others described only a partial softening of preconceptions \punit{EP1, EP2, EP3, EP5, EP9, CP7}, and still others reported no change \punit{EP6, EP7, CP2, CP4, CP6}. Across both conditions, participants who reported positive attitude change described how prior stereotypical assumptions were corrected through exposure to the outgroup's authentic viewpoints.
\begin{quote}
    \textit{``Since I often see Chinese tourists, I used to feel that they weren't very mindful of their surroundings. However, I felt that Chinese international students seemed quite self-aware, recognizing a thoughtfulness about them and came to respect them more than before.''} \punit{EP4}
\end{quote}

However, de-stereotyping did not necessarily lead to positive attitude change. In the experimental condition, some participants \punit{EP1, EP7, EP9} cognitively categorized AI as fundamentally distinct from humans, which constrained attitude change. 
\begin{quote}
    \textit{``My impression did shift somewhat for the better. But since this was ultimately an AI, not the actual person, my preconceptions just softened a little.''} \punit{EP9}
\end{quote}



\subsubsection{Lack of Generalization to Chinese people in general}

Although some participants reported de-stereotyping toward Chinese international students, only two participants \punit{EP5, CP1} reported attitude change toward the Chinese people in general. Participants differentiated Chinese international students from the Chinese people in general, refraining from generalizing their specific experiences to the entire group.
\begin{quote}
    \textit{``International students come with the intention of learning about Japan. But when it comes to Chinese people in general, it’s not always because they like Japan.''} \punit{EP9}
\end{quote}

\subsection{Psychological Changes for Future Contact}

\subsubsection{Reduction in Intergroup Anxiety}

The two conditions differed in reported changes in intergroup anxiety. While a reduction in intergroup anxiety was consistently reported in the experimental condition \punit{EP1, EP3, EP4, EP5, EP7, EP9}, only two participants \punit{CP1, CP6} described such a reduction in the control condition.

Participants in the experimental condition attributed this reduction to several factors: a recognition of shared values and orientations between the two groups \punit{EP1}, the dissolution of prior misconceptions about the outgroup \punit{EP3}, and the experience of constructive dialogue through GroupEnvoy \punit{EP3, EP4}.
\begin{quote}
\textit{``AI’s arguments made me realize that outgroup members were much calmer and more constructive than I had expected. This has alleviated my initial anxiety and reluctance regarding discussing such topics.''} \punit{EP4}
\end{quote}

In the control condition, CP1 attributed reduced anxiety to recognizing the outgroup's sincerity through the records. CP6 noted that while anxiety about direct contact remained high, this discussion would lower it compared to having no prior contact.

Despite these reductions, participants who experienced anxiety reduction reported that concerns specific to direct contact remained. Two sources of residual anxiety were shared across both conditions: a fear of offending due to limited cultural knowledge \punit{EP8, CP2}, and anticipated difficulty in communication due to language and cultural differences \punit{EP2, EP7, EP9, CP3, CP5}.

\subsubsection{Improvement in Outcome Expectancies}

Across both conditions, participants attributed improved outcome expectancies primarily to confirming the outgroup's cooperative orientation. In the experimental condition, participants \punit{EP1, EP3, EP4, EP6} attributed improved outcome expectancies to the experience of constructive dialogue via GroupEnvoy.
\begin{quote}
    \textit{``I learned that sharing clear goals leads to productive discussions. This makes me feel ready to talk things through, even with actual Chinese international students.''} \punit{EP1}
\end{quote}

Similarly, in the control condition, participants \punit{CP3, CP5, CP6} reported improved outcome expectancies grounded in the outgroup's cooperative attitudes and interest in Japanese culture, as reflected in the discussion transcripts.
\begin{quote}
    \textit{``Chinese participants were surprisingly cooperative and enthusiastic, which made collective events seem highly achievable and promising.''} \punit{CP5}
\end{quote}

However, in the experimental condition, this improvement was constrained by concerns that GroupEnvoy's responses might be overly accommodating rather than reflecting genuine outgroup perspectives \punit{EP5, EP9}. In the control condition, some participants \punit{CP1, CP3, CP8} argued that written records alone were insufficient to estimate how a live discussion would unfold, and expressed a preference for direct dialogue. CP8 noted that \textit{``the written information is all there is, so anything beyond that can only be inferred.''} Some other participants \punit{CP2, CP4} reported no improvement in outcome expectancies due to the anticipated difficulties stemming from language and cultural differences.

\subsubsection{Future Contact Intentions}

The two conditions differed qualitatively in how the dialogue shaped participants' future contact intentions. In the experimental condition, the predominant change was a reduction in intergroup anxiety rather than an increase in future contact intentions. While participants \punit{EP1, EP3, EP5, EP6, EP7, EP8} reported that their resistance to hypothetical direct interaction had decreased, their spontaneous desire to seek such contact remained unchanged, noting that proactive contact would require a shared purpose or topic.
\begin{quote}
    \textit{``My active side towards interaction has not changed at all, but the passive side has changed a little more positively.''} \punit{EP5}
\end{quote}

In the control condition, participants \punit{CP1, CP3, CP5, CP7} reported increased future contact intentions, driven by heightened interest in and familiarity with the outgroup.
\begin{quote}
    \textit{``Since I learned that Chinese international students have a positive impression of Japan, we could have enjoyable interactions without putting up too many barriers. After the experiment, I felt more motivated to interact.''} \punit{CP3}
\end{quote}
However, some participants \punit{CP4, CP6, CP8} in the control condition reported no change because their motivation to interact had already been high before the study.

Taken together, these findings point to divergent pathways: participants in the experimental condition primarily experienced a reduction in intergroup anxiety, whereas those in the control condition described increased motivation for future contact.

\subsection{Outgroup Review of GroupEnvoy's Responses} \label{sec:outgroup_review_responses}

Both reviewers classified the GroupEnvoy's responses as either explicitly stated or aligned with outgroup views (P1: 29/30; P2: 28/30), suggesting that the agent largely preserved the content and perspective of the original outgroup discussion. The small number of discrepancies revealed two distinct failure modes: factual confusion between outgroup members' biographical details (i.e., \textit{persona cross-contamination}) and the generation of content not grounded in the source discussion (i.e., \textit{unattested inferences}. These limitations are examined further in the Discussion (Section~\ref{sec:dr2_discussion}).

\section{Discussion}

\subsection{RQ1: Psychological Effects of AI-Mediated Intergroup Contact}

To address RQ1, this section examines psychological changes within and between conditions. Quantitatively, significant main effects of Time across multiple measures indicate that exposure to outgroup perspectives promoted psychological change regardless of modality. Although no Group $\times$ Time interaction reached significance, medium-to-large effect sizes for intergroup anxiety and perspective-taking suggest a directional advantage of AI-mediated contact. Qualitative results further illuminate the underlying mechanisms.

\subsubsection{Interactivity as the Critical Mechanism: Perspective-Taking and Anxiety Reduction}

Although the results did not reach statistical significance, our qualitative findings and medium-to-large effect sizes suggest that AI-mediated contact with GroupEnvoy offers distinct advantages over passive text exposure in reducing intergroup anxiety and enhancing perspective-taking. The observed patterns suggest that the two modalities may serve different intervention goals: text-based transcripts may be better suited to promoting outgroup understanding through sustained perspective-taking, whereas AI-mediated contact may more effectively reduce intergroup anxiety, thereby lowering the psychological barriers to future direct contact. 

Additionally, as GroupEnvoy required participants to actively formulate responses to outgroup views, participants engaged in an iterative dialogue that facilitated self-other merging, a psychological process in which the boundary between self and other blurs, thereby reducing psychological distance \cite{bruneau_power_2012}. GroupEnvoy's asynchronous, text-based format afforded time for deliberation before each exchange, enhancing self-presentational control. Prior E-contact research indicates that this flexibility builds participants' confidence in making a desired impression, thereby enhancing outcome expectancies, which in turn reduce intergroup anxiety \cite{white_improving_2019, plant_responses_2004}. Consistent with this, outcome expectancies improved significantly over time across both conditions. Future research should investigate whether outcome expectancies specifically mediate anxiety reduction in AI-mediated contact and how such \textit{interactivity} shapes these expectancies.

However, there are some confounding factors around \textit{interactivity}. The novelty of the interface may have heightened attentiveness in the experimental condition, and the facilitation role introduced asymmetry in group dynamics. In the control condition, the time required to read the materials before discussion may have reduced the available time for discussion.

\subsubsection{Divergent Pathways: Intergroup Anxiety and Future Contact Intentions} \label{sec:divergent_pathways}

The two conditions qualitatively diverged in distinct ways. While the experimental condition largely demonstrated reduced intergroup anxiety alongside stable future contact intentions, the control condition suggested increased future contact intentions despite weaker anxiety reduction. This divergence suggests that contact avoidance and approach motivation operate independently rather than as endpoints of a single continuum. \citet{greenland_intergroup_2012} formalize this distinction, identifying self-anxiety and other-anxiety as two independent dimensions. Other-anxiety is defined as concerns about how the outgroup will behave toward the self, while self-anxiety is defined as concerns about appearing prejudiced, which predicts both approach and avoidance tendencies. \citet{paolini_self-expansion_2016} further shows that approach motivation stems from self-expansion expectancies and operates independently of avoidance reduction.

Consistent with this framework, GroupEnvoy primarily reduced other-anxiety. Its asynchronous, text-based format and predictable responses mitigated threat appraisals \cite{trawalter_predicting_2009} and reduced intergroup anxiety without fostering future contact intentions. In contrast, exposure to the outgroup's authentic viewpoints through written transcripts likely activated self-expansion expectancies because participants recognized that dissimilar others offer novel resources and perspectives \cite{paolini_self-expansion_2016}. Its explicit identity as an AI agent likely stifled this effect, as participants could not readily form self-expansion expectancies toward a conversational agent. These divergent pathways suggest that AI-mediated and text-based formats differ in kind rather than degree. However, if AI-mediated contact reduces intergroup anxiety without motivating direct contact, it risks becoming a substitute rather than a preparatory step. Future implementations should therefore incorporate explicit transitions from AI-mediated to direct contact.

\subsection{RQ2: Design Implications for AI-Mediated Intergroup Contact} \label{sec:design_implications}

This section evaluates GroupEnvoy against its design rationales to derive design implications for AI-mediated intergroup contact. Table \ref{tab:design_implications} summarizes the key findings and revised design rationales.

\begin{table*}[t]
\caption{Summary of design rationale evaluation: findings and revised design rationales for AI-mediated intergroup contact.}
\label{tab:design_implications}
\begin{tabular}{>{\raggedright\arraybackslash}p{0.05\textwidth} >{\raggedright\arraybackslash}p{0.11\textwidth} >{\raggedright\arraybackslash}p{0.38\textwidth} >{\raggedright\arraybackslash}p{0.35\textwidth}}
\toprule
\textbf{DR} & \textbf{Sub-element} & \textbf{Finding} & \textbf{Revised Design Rationales} \\
\midrule
\multirow{2}{*}{DR1}
  & Equal Status & Excessive accommodation can be perceived as sycophancy. & Balancing responsiveness with firmness to maintain outgroup distinctiveness. \\
  & Cooperation & Proactive AI facilitation may reduce participant initiative. & Constraining facilitative roles to preserve participant ownership of the dialogue. \\
\midrule
\multirow{3}{*}{DR2}
  & Typicality & Improved outgroup attitudes may not generalize due to subtyping. & Including diverse subgroups to broaden the scope of outgroup representation. \\
  & Individuation & Persona-level familiarity fades as collective viewpoints dominate. &  Reintroducing individual narratives periodically to anchor the collective perspectives. \\
  & Authenticity & Persona cross-contamination and unattested inferences may undermine response authenticity. & Implementing retrieval-augmented architectures and strict source-grounded generation.\\
\midrule
\multirow{2}{*}{DR3}
  & Affective Empathy & AI-mediated contact elicits meta-perceptions of outgroup evaluations of the ingroup, rather than compassion for outgroup experiences. & Distinguishing empathy target: compassion for outgroup, or meta-perception of outgroup evaluations of the ingroup. \\
  & Perspective-taking & Users may delegate perspective-taking effort to the AI agent. & Scaffolding engagement by prompting user inference before AI response. \\
\bottomrule
\end{tabular}
\end{table*}

\subsubsection{DR1: Equal Status and Cooperation}

Participants broadly perceived \textit{equal status} in the experimental condition. By consistently articulating outgroup values, the agent positioned itself as an equal counterpart rather than a subordinate tool. However, the accommodation response style occasionally undermined \textit{equal status}, revealing that the boundary between responsiveness and sycophancy is highly subjective. Therefore, achieving \textit{equal status} demands finely calibrated responsiveness that validates ingroup perspectives without compromising those of the outgroup.

A sense of \textit{cooperation} emerged largely in the experimental condition. However, GroupEnvoy inadvertently reduced participants' initiative to drive the discussion forward by assuming a facilitative role in structuring the dialogue. Since this tendency contradicts Allport's conception of genuine cooperation \cite{pettigrew_intergroup_1998}, future designs should constrain the agent's facilitative functions to preserve participant ownership of the dialogue.

\subsubsection{DR2: Faithfully Conveying Outgroup Perspectives} \label{sec:dr2_discussion}

A sense of \textit{typicality} was established as participants in the experimental condition perceived GroupEnvoy as representing Chinese international students, suggesting that the agent's collective framing successfully anchored group salience during the interaction. However, improved attitudes toward Chinese international students did not generalize to the Chinese people in general. Subtyping theory \cite{johnston_cognitive_1992, wilder_intergroup_1984} accounts for this pattern: participants perceived Chinese international students as a self-selected group who had chosen to come to Japan, thereby treating them as distinct from other Chinese people. Generalization thus stopped at the subgroup boundary, as the contact partner was not seen as representative of the broader outgroup \cite{pettigrew_intergroup_1998, hewstone_contact_1986}. This limitation is not specific to AI-mediated contact but reflects a broader challenge in intergroup contact research \cite{brown_integrative_2005}. Practitioners deploying this approach should ensure that the demographics of the represented group (e.g., age, gender, occupation) align with those of the intended outgroup and are perceptible through the agent's responses, while ensuring that participants clearly understand the agent's responses structurally reflect specific outgroup views.

\textit{Individuation} was initially supported by persona data, which fostered familiarity and a sense of human-likeness among participants. However, as the dialogue progressed, GroupEnvoy increasingly spoke on behalf of the group as a whole, with persona-level details receding in favor of arguments representing collective outgroup viewpoints. This attenuation reflects a structural tension between \textit{individuation} and \textit{typicality} inherent to AI-mediated contact.

Evaluations of \textit{authenticity} demonstrate that grounding an LLM agent in outgroup data successfully produces outputs recognized as faithful to the authentic viewpoints of outgroup members, mitigating the perceived lack of \textit{authenticity} typically associated with generative AI \cite{chiang_enhancing_2024, lee_amplifying_2025}. However, the model's own training biases may reproduce or reinforce stereotypes \cite{hu_2024_Generative}. The outgroup review in Section \ref{sec:outgroup_review_responses} further revealed two failure modes, \textit{persona cross-contamination} and \textit{unattested inferences}. To mitigate these risks, future systems require architectural safeguards beyond prompt design, such as (1) \textit{retrieval-augmented architectures} that partition each persona's data into isolated context blocks with explicit speaker attribution; (2) \textit{source-grounded generation} that constrains personal narratives to verifiable excerpts.

\subsubsection{DR3: Perspective-Taking and Affective Empathy}

In terms of \textit{affective empathy}, participants in the experimental and control conditions underwent qualitatively different empathic processes. While participants in the control condition frequently expressed compassion for the outgroup's lived experiences, participants in the experimental condition predominantly engaged with \textit{meta-perceptions}, referring to inferences about how the outgroup evaluates the ingroup \cite{galinsky_perspective-taking_2005, frey_being_2006, wout_when_2010}. Participants in the control condition projected familiar emotions onto the outgroup by encountering accounts of the outgroup's lived experiences in Japan. \citet{galinsky_perspective-taking_2005} describes this as \textit{application of the self to the other}, a process that has been shown to reduce stereotyping. 

In contrast, participants in the experimental condition continuously engaged with GroupEnvoy's evaluations of Japanese students, exposing them to the concern of being negatively evaluated by the outgroup as stereotypical group members rather than as individuals \cite{frey_being_2006}. Such anticipated negative evaluation constitutes a primary source of intergroup anxiety \cite{stephan_intergroup_1985}. \citet{turner_test_2008} identify \textit{positive outgroup norms}, the belief that the outgroup seeks positive relations with the ingroup, as a key mechanism for reducing intergroup anxiety. When participants encountered GroupEnvoy’s evaluations of Japanese students as largely positive and consistent with ingroup self-perceptions, they came to perceive \textit{positive outgroup norms}, thereby reducing the concern of negative outgroup evaluation. This highlights the importance of creating settings in which the ingroup can anticipate positive future contact with the outgroup, ultimately fostering intergroup relations.

\textit{Perspective-taking} improved significantly over time across both conditions. Although the $\text{Group} \times \text{Time}$ interaction did not reach statistical significance, it demonstrated a medium-to-large effect size. Qualitative findings suggest that AI-mediated contact prompted some participants to delegate perspective-taking to GroupEnvoy, thereby diminishing their effortful cognitive engagement with outgroup viewpoints. Since perspective-taking reduces prejudice through effortful cognitive engagement \cite{davis_empathy_1994, broockman_durably_2016}, delegating this process to GroupEnvoy risks producing shallower and less persistent attitude change. This highlights the importance of agents scaffolding users' perspective-taking efforts rather than substituting for them, particularly given that sustained reliance on LLMs reduces independent cognitive engagement and performance over time \cite{kosmyna_2025_your}. Future work should therefore examine whether AI-mediated contact fosters genuine outgroup understanding through perspective-taking or inadvertently substitutes the inferential effort.

\subsection{Design Implications for Conversational User Interface} \label{sec:cui_implications}

When a conversational agent speaks on behalf of real people rather than assisting the user directly, two design considerations become critical. First, while responsiveness to the user is typically well received, it can be perceived as sychophancy that undermines fidelity to the represented party. Despite GroupEnvoy being grounded in empirical outgroup data and programmed to disagree with conflicting user proposals, some participants still perceived it as sycophantic. This indicates that perceived sycophancy arises as much from conversational tone and interaction style as from factual grounding. Designers should therefore calibrate how the agent communicates, not just what it knows, to ensure the agent is perceived as an equal counterpart rather than a compliant assistant. 

Second, proactively presenting a third party's perspectives risks substituting rather than scaffolding users' cognition. GroupEnvoy exhibited this \textit{delegation effect}, with some participants offloading perspective-taking to the agent, even though it continuously posed questions from outgroup perspectives. To achieve scaffolding, interaction should elicit users' own inferences before presenting its perspective, preventing users from bypassing the inferential effort.

\subsection{Limitations and Future Work} \label{sec:limitations}

Several statistical limitations temper the interpretation of our findings. The small sample size (N = 17; n = 9 Experimental, n = 8 Control) constrained statistical power, and no Group $\times$ Time interaction reached significance. The medium-to-large effect sizes observed for intergroup anxiety ($\eta_p^2 = .158$) and perspective-taking ($\eta_p^2 = .114$) suggest that these null results reflect power limitations rather than the absence of effect. Given that participants were nested within triads and facilitators had a distinct experience from other participants, future work would benefit from multilevel modeling to address potential dependencies and role-specific effects, alongside larger-scale replications to confirm the observed trends. Furthermore, future work should assess baseline attitudes toward AI, given participants' tendency to categorize AI as fundamentally distinct from humans (Section \ref{sec:de_stereotyping}).

Beyond statistical constraints, our experimental design introduces additional limitations. Our findings are situated within a specific intergroup context, leaving open the question of whether AI-mediated contact produces comparable effects across racial, religious, or political divisions. Our design also imposed three structural constraints: a unilateral setup due to recruitment limitations, a single post-session measurement timepoint, and confounded \textit{interactivity}. Future work should employ a bilateral design to investigate the psychological differences in how dominant and non-dominant groups respond to AI-mediated contact, incorporate longitudinal follow-ups to reveal whether multiple sessions yield cumulative benefits \cite{pettigrew_intergroup_1998}, and include additional conditions, such as presenting GroupEnvoy's pre-generated responses as a static document, to isolate the contribution of \textit{interactivity} per se.

A further limitation concerns the structural tension between \textit{typicality} and \textit{individuation}. Future designs should periodically reintroduce persona-level information during the dialogue, anchoring collective views to individuals without compromising group salience. A comparison between GroupEnvoy's collective representation and a more individuated, persona-based approach would further clarify the distinct roles of \textit{typicality} in mitigating subtyping and individuation in fostering outgroup familiarity.

\section{Conclusion}
This study introduced AI-mediated contact, a paradigm in which a conversational agent grounded in outgroup discussion data conveys outgroup perspectives during ingroup discussions, and presented GroupEnvoy as its implementation. Although no Group $\times$ Time interaction was significant, medium-to-large effect sizes for intergroup anxiety and perspective-taking suggest that interactive AI-mediated contact may outperform passive text exposure. Qualitative analysis demonstrated distinct mechanisms in response to experimental conditions, including enhanced outcome expectancies, engagement with \textit{meta-perceptions} regarding the outgroup's perception of the ingroup, and three distinct patterns of perspective-taking. Our findings also indicate that effective AI-mediated contact requires calibrating agent responsiveness to prevent sycophancy, sustaining \textit{individuation} within collective outgroup representation, and scaffolding rather than substituting users' perspective-taking effort. These findings position AI-mediated contact as a promising preparatory step toward direct intergroup contact, with future work needed to validate these effects on a larger scale and across diverse intergroup contexts.

\begin{acks}
We used Claude to improve the clarity and grammar of our writing and to clarify the argumentative structure of previous research. We also consulted Claude's suggestions on potential segment boundaries and candidate codes for post-experiment interview transcripts, but all segmentation and code assignment decisions reported in this paper were made independently by the authors.

This research was supported by The Nippon Foundation HUMAI Program and is part of the results of Value Exchange Engineering, a joint research project between Mercari R4D Lab and RIISE (Research Institute for an Inclusive Society through Engineering).
\end{acks}

\bibliographystyle{ACM-Reference-Format}
\bibliography{references}

\end{document}